\documentstyle[aps,eqsecnum,preprint]{revtex}
\begin{document}
\draft
\title{\bf Mathematical Foundations of Regular Quantum Graphs}
\author{R. Bl\"umel, Yu. Dabaghian and R. V. Jensen}
\address{Department of Physics, Wesleyan University,
Middletown, CT 06459-0155}
\date{\today}
\maketitle

\begin{abstract}
We define a class of quantum systems called regular
quantum graphs.
Although their dynamics is chaotic in the
classical limit with positive topological entropy,
the spectrum of regular quantum graphs
is explicitly computable analytically and exactly,
state by state,
by means of periodic orbit expansions.
We prove analytically
that the
periodic orbit series exist and
converge to the correct spectral eigenvalues.
We investigate the convergence properties of
the periodic orbit series and prove rigorously that
both conditionally convergent and absolutely convergent
cases can be found. We compare the periodic orbit
expansion technique with Lagrange's inversion formula.
While both methods work and yield exact results,
the periodic orbit expansion technique has conceptual value
since all the terms in the expansion have direct physical
meaning and
higher order corrections are obtained according to
physically obvious rules.
In addition
our periodic orbit expansions
provide explicit analytical solutions
for many classic text-book examples of quantum mechanics
that previously could only
be solved using graphical or numerical techniques.
\end{abstract}


\section{Introduction}
At a first glance
it may seem surprising that
many chaotic dynamical systems have explicit analytical
solutions. But many examples are readily at hand.
The shift map \cite{Devaney,Ott}
\begin{equation}
   x_{n+1}=(2x_n)\ \ {\rm mod}\ 1,\ \ x_n\in{\rm{\bf R}},
   \ \ n=0,1,2,\ldots ,
\label{shiftmap}
\end{equation}
for instance, is ``Bernoulli'' \cite{Ott},
the strongest
form of chaos. Nevertheless the shift map is readily solved
explicitly,
\begin{equation}
     x_n=(2^n\, x_0)\ \ {\rm mod}\ 1,\ \ x_n\in{\rm{\bf R}},
     \ \ n=0,1,2,\ldots .
\label{shiftsol}
\end{equation}
Another example is provided
by the logistic mapping
\begin{equation}
  x_{n+1}=\mu x_n(1-x_n),\ \ x_n\in [0,1],\ \
  0\leq \mu\leq 4,
  \ \ n=0,1,2,\ldots ,
\label{logisticmap}
\end{equation}
widely used in population dynamics
\cite{Devaney,Ott,May}.
For $\mu=4$ this mapping is equivalent with the shift map
\cite{UlNeu}
and therefore again completely chaotic.
Yet an explicit
solution, valid at $\mu=4$, is given by\cite{UlNeu,BR}:
\begin{equation}
x_n=\sin^2\left(2^n\arcsin\sqrt{x_0}\,\right),\ \ x_0\in [0,1].
\label{logisol}
\end{equation}
Therefore, as far as classical chaos is concerned, there is
no basis for the belief that classically chaotic systems do
not allow for explicit analytical solutions. But what about the
quantized versions of classically chaotic systems,
commonly known as quantum chaotic
systems\cite{Gutz,STOECK}?
Here, too, the answer is affirmative. It was shown recently
\cite{PRL,PRE1,PRE2,JETPL,JETP} that
regular quantum graphs provide
a large class of
explicitly solvable quantum chaotic systems.
In order to strengthen first analytical and
numerical results presented in
\cite{PRL,PRE1,PRE2,JETPL,JETP},
it is the purpose of this paper to show that
the explicit periodic orbit
expansions obtained in \cite{PRL,PRE1,PRE2,JETPL,JETP}
are more than formal identities. We will
prove below that (i) the spectrum of regular quantum graphs
is computable explicitly and
analytically, state by state, via
convergent periodic orbit expansions and (ii)
the periodic orbit series converge to the correct spectral
eigenvalues.

The main body of this paper is organized as follows.
In Sec.~II we summarize the basics of quantum graph theory
and derive the general form of the spectral equation.
In Sec.~III we define regular quantum graphs. In Sec.~IV we
present the explicit periodic orbit expansions of
individual eigenvalues of regular quantum graphs.
We also specify a summation scheme that guarantees convergence
to the correct results.
The derivations presented in
Sec.~IV are mathematically rigorous except for one
step where we interchange integration and summation to
arrive at our final results. This is a
``dangerous operation'',
which is not allowed without further investigation.
This point is resolved in
Sec.~V,
where we present the analytical proofs that
justify the interchange of integration and summation performed
in Sec.~IV.
This result establishes that the
periodic orbit expansions
investigated in this paper
converge in the usual sense of
elementary analysis.
In Sec.~VI we investigate the convergence properties
of the periodic orbit series obtained in Sec.~IV.
We prove analytically that there exists at least
one quantum graph for which the convergence
is not absolute, but conditional.
According to Riemann's well-known reordering theorem
(see, e.g. \cite{EL}, volume II,
page 33), it is possible to reorder the terms
of a conditionally convergent sum in such a way
that it converges to any prescribed number.
This is the reason why
in Sec.~IV we place so much emphasis on specifying
a summation scheme that
guarantees convergence of the periodic orbit sums
to the correct spectral eigenvalues.
In Sec.~VII we present an alternative way of solving
the spectral equations of regular quantum graphs:
Lagrange's inversion formula. Although a perfectly
sound and fast-converging method for solving the
spectral equation, it lacks the physical appeal
of periodic orbit expansions which are based
on concrete physical objects such as periodic orbits
and their geometrical and dynamical properties.
In Sec.~VIII we discuss our results, point out
promising research directions and
conclude the paper.

Since many scientists will find the existence of convergent
periodic orbit expansions
surprising, we found it necessary to establish this result with
an extra degree of rigor.
This necessarily requires strong, but somewhat technical
proofs. However,
in order not to break the flow of the paper, we
adopt a hierarchical approach
presenting the material in three stages.
The most pertinent aspects
of our results are presented in the main text
as outlined above.
Supporting
higher-level, but formal, material is presented
in Appendix A.
Lower level material, such as
formulae, definitions and lemmas, is relegated to
Appendix B.

In order for the proofs to be convincing,
and to be accessible to a wide readership,
we used only concepts of elementary
undergraduate analysis in our proofs, altogether avoiding
advanced mathematical concepts, such as distributions
\cite{DEMI,WALT}.
This is an important point since
our paper seeks
mathematical rigor on a common basis acceptable
to all readers.
In this spirit the alert reader will notice that we
completely avoid the use of Dirac's
delta ``function'' \cite{Dirac}.
This is necessary since the delta ``function'' is
a distribution
\cite{DEMI,WALT},
a concept we found to be highly
confusing for the general reader.
Although there is nothing ``wrong'' with the
delta ``function'', if treated properly as
a distribution or linear functional,
the confusion surrounding the delta ``function''
started with von Neumann's critique \cite{vNeumann}
at a time when the modern tools of distribution theory
were not yet available. Therefore, although
ultimately completely
unjustified, von Neumann's critique \cite{vNeumann}
tainted the delta ``function'' to such an extent that
it still can't be used in a rigorous context without
causing heated debate. Thus, for the sake of simplicity
and clarity of our arguments, we prefer to avoid it.
As a consequence, the
presentation of the material in this paper
and all of our proofs are conducted
without using the concept of level densities, which
are usually defined with the help of Dirac's delta ``function''.

We emphasize that there is a fundamental difference between
an analytical solution and an
{\it explicit} analytical solution. While the spectral equation
for quantum graphs is known in great detail \cite{QGT1},
and periodic orbit expansions for the spectral density and
the staircase function of quantum graphs are known \cite{QGT1},
these results are all implicit. This means that they do not
yield the spectral eigenvalues in the form ``$E_n=\ldots$''.
It is the novelty of this paper to obtain explicit analytical
expressions of this type for a wide class of
quantum chaotic systems, and prove their validity
and convergence with mathematical rigor.

\section{Quantum Graphs and Spectral Equation}
The properties of the Laplacian operator
on graphs have been studied
in great detail in the mathematical literature \cite{Novi1,Novi2}
and the study of quantum mechanics on graphs
has attracted
considerable attention among
chemists and physicists
(see, e.g., \cite{CHEM,ACD,USP} and references therein),
especially
in the quantum chaos community
\cite{QGT1,ACD,QGT2,paradigm,BG1,PZK}.
The purpose of this section is to acquaint the reader
with the main ideas of quantum graph theory.
Since many excellent
publications on the theory of quantum graphs
are available (see, e.g.,
\cite{QGT1,ACD,USP,QGT2,paradigm,BG1,PZK}),
we will present only those ideas and
facts that are of direct relevance to the subject of this paper.

A quantum graph consists of a network of bonds and vertices
with a quantum particle travelling on it. An example of
a graph with ten bonds and six vertices is shown in Fig.~1.
The number of bonds is denoted by $N_B$, the number of vertices
by $N_V$.
In this paper we focus entirely on finite
quantum graphs, i.e.
$N_B$, $N_V<\infty$.
We define directed bonds on the graph
such that the bond $B_{ij}$ connecting vertex number $i$
with vertex number $j$
is different from the bond $B_{ji}$ connecting
the vertices in the opposite direction.
There are $2N_B$
directed bonds. It is useful to define the linearized bond
index $\Lambda=1,\ldots,N_B$:
$(ij)\mapsto\Lambda$, $i<j$, $i$ and $j$ connected.
The index $\Lambda$ labels sequentially all $N_B$ directed
bonds of the
graph with $i<j$. For the directed bonds $B_{ij}$ with $i>j$
we define $\Lambda(ij)=-\Lambda(ji)$. This way
the sign of the
counting index reflects the directionality of the bond.
The network of bonds and vertices
defines the graph's topology. The topology of a graph alone
does not completely specify the quantum graph.
This is so because, for instance,
the bonds of the graph may be dressed with potentials \cite{PRE1}.
Since the quantum graphs we study in this paper are
finite, bounded systems, their spectra
are discrete.

The spectrum of a quantum graph
is obtained by solving the one-dimensional
Schr\"odinger equation on the graph subjected to the
usual boundary conditions of
continuity and quantum flux conservation.
A particularly useful way of obtaining the spectral
equation for a given quantum graph is the scattering
quantization approach \cite{QGT1,QGT2,SQA} which yields the
spectral equation in the form
\begin{equation}
\det[1-S(k)]=0,
\label{specdet}
\end{equation}
where $S$ is the quantum scattering matrix of the graph
\cite{QGT1,ACD,QGT2}
and $k$ is the wave number, related
to the energy via $E=k^2$.
For our purposes it is sufficient
to know that the $S$ matrix is of dimension $2N_B\times 2N_B$ and
can be decomposed into \cite{JETP,QGT1}
\begin{equation}
S(k) =  D(k)\, U(k),
\label{decomp1}
\end{equation}
where $U(k)$ is a $2N_B\times 2N_B$ unitary matrix and
$D(k)$ is a diagonal matrix of the form
\begin{equation}
D_{\Lambda,\Lambda'}(k) =
\exp\left[iL_{\Lambda}(k)\right]\ \delta_{\Lambda,\Lambda'},
\ \ \Lambda,\Lambda'=\pm 1,\ldots,\pm N_B
\label{DD}
\end{equation}
with $L_{\Lambda}(k)\in {\rm{\bf R}}$,
$L_{\Lambda}(k)=L_{-\Lambda}(k)>0$.
The ordering of the matrices $D$ and $U$ in (\ref{decomp1})
is neither unique nor important in the present context.
It depends on the details of how ``in'' and ``out''
channels are assigned to the indices of the $S$ matrix.
Since the computation of the energy spectrum
involves only traces and determinants, the precise
ordering of $D$ and $U$ in (\ref{decomp1}) does not
affect our final results.

Physically the quantities $L_{\Lambda}(k)$ are the
time-reversal invariant parts of the bond actions
\cite{PRE2,JETP,QGT1}. A possible time-reversal
breaking part of the bond actions is
understood to be absorbed
in the matrix $U$ \cite{JETP}.
For simplicity we will in the following refer
to $L_{\Lambda}$ as the bond action of the bond
$B_{\Lambda}$.

In this paper we focus exclusively on scaling quantum graphs
\cite{PRL,PRE1,PRE2,JETPL,JETP,JMP}.
In this case
the matrix $U$ is a constant
matrix, independent of $k$, and the
actions $L_{\Lambda}(k)$ in
(\ref{DD}) split into the product
$L_{\Lambda}(k) =  L_{\Lambda}^{(0)}k$,
where $L_{\Lambda}^{(0)}\in {\rm{\bf R}}$
is a constant, the reduced bond action of the bond
$B_{\Lambda}$.
Physically the scaling case is an important
case since it describes systems free from
phase-space metamorphoses \cite{morph}.
Scaling systems of this type
arise in many physical contexts, for instance
in microwave cavities partially filled with a dielectric
\cite{SKB,Bauch,Found},
or Rydberg atoms in external fields \cite{Tom}.

It is possible to write (\ref{specdet}) as a linear
combination of trigonometric functions whose frequencies
are directly related to the bond actions. In order to
derive this representation
we start by noting that
\begin{equation}
\det[1-S]=\det\left(S^{1/2}\right)\,
(-4)^{N_B}\, \prod_{l=1}^{2N_B}\, \sin(\sigma_l/2),
\label{dstruc}
\end{equation}
where
$\sigma_l\in {\rm{\bf R}}$
are the eigenphases of the unitary
matrix $S$.
>From (\ref{dstruc}) we obtain the important result that
\begin{equation}
\det\left(S^{-1/2}\right)\det[1-S]\, \in\, {\rm{\bf R}}.
\label{real1}
\end{equation}
According to
(\ref{decomp1}) we have
\begin{equation}
\det(S^{-1/2})=\exp(-i\tau/2)\,
\exp\left(-i\sum_{\Lambda=1}^{N_B}L^{(0)}_{\Lambda}k\right),
\label{det1}
\end{equation}
where $\tau$ is the eigenphase of the unitary matrix $U$,
i.e. $\det(U)=\exp(i\tau)$.
Using the
decomposition (\ref{decomp1}) of the $S$ matrix
we write the spectral determinant (\ref{specdet})
in the form
\begin{equation}
\det\left[(-U^{-1})+D\right]=0,
\label{specdet1}
\end{equation}
so that we can directly apply the results in
Ref. \cite{Aitken}.
According to Ref. \cite{Aitken}, p. 87, the determinant
(\ref{specdet1}) is a polynomial in the $2N_B$ variables
$\exp(iL^{(0)}_{\Lambda}k)$,
$\Lambda=\pm 1,\ldots,\pm N_B$, whose
coefficients are the principal sub-determinants
\cite{Aitken}
of $-U^{-1}$.
Using
$L^{(0)}_{-\Lambda}=
L^{(0)}_{\Lambda}$
together with
the fact that $\det(U^{-1})$ is the principal
sub-determinant of $-U^{-1}$ of order zero
\cite{Aitken},
we obtain that (\ref{specdet1}) is of the form
\begin{equation}
\det\left[(-U^{-1})+D\right]= \det(U^{-1}) +
\sum_{n=1}^{N_B}\sum_{i_1,\ldots,i_n}\sum_{\alpha^{(n)}}\,
A_{n;i_1,\ldots, i_n;\alpha^{(n)}}\,
\exp\left[i\sum_{\Lambda\in\{i_1,\ldots,i_n\}}
\alpha^{(n)}_{\Lambda}L_{\Lambda}^{(0)}k\right],
\label{xyz}
\end{equation}
where
$1\leq i_1 <
i_2<\ldots < i_n\leq N_B$,
$\alpha^{(n)}$ is an integer
array of length $n$ containing
a ``1,2-pattern'', i.e.
$\alpha^{(n)}_{\Lambda}\in\{1,2\}$,
and
$A_{n;i_1,\ldots, i_n;\alpha^{(n)}}$
are complex coefficients that can be computed
from the principal sub-determinants of
$-U^{-1}$.
Because of (\ref{real1}), we have
\begin{equation}
\det\left( S^{-1/2}\right)\, \det(U)\, \det\left(-U^{-1}+D
\right)\ \in \ {\rm{\bf R}}.
\label{real2}
\end{equation}
Defining
\begin{equation}
\omega_0=\sum_{\Lambda=1}^{N_B}\, L^{(0)}_{\Lambda}
\label{omega0}
\end{equation}
and using (\ref{det1}) and (\ref{xyz}),
we obtain (\ref{real2}) in the form
\begin{equation}
\exp[-i(\omega_0 k+\tau/2)]+
\sum_{n=1}^{N_B}\sum_{i_1,\ldots,i_n}\sum_{\alpha^{(n)}}\,
A_{n;i_1,\ldots, i_n;\alpha^{(n)}}\,
\exp\left\{-i[\beta_{n;i_1,\ldots,i_n;\alpha^{(n)}}k
-\tau/2]\right\}
\ \ \in\ {\rm{\bf R}},
\label{real3}
\end{equation}
where
\begin{equation}
\beta_{n;i_1,\ldots,i_n;\alpha^{(n)}} =
\omega_0-
\sum_{\Lambda\in\{i_1,\ldots,i_n\}}
\alpha^{(n)}_{\Lambda}L_{\Lambda}^{(0)}.
\label{betdef}
\end{equation}
We define the frequencies
\begin{equation}
\omega_{n;i_1,\ldots,i_n;\alpha^{(n)}}=
|\beta_{n;i_1,\ldots,i_n;\alpha^{(n)}}|.
\label{omdef}
\end{equation}
Because of $L_{\Lambda}^{(0)}>0$ for all $\Lambda$
and the structure of $\alpha^{(n)}$,
the largest frequency in (\ref{omdef}) is
$\omega_0$ defined in (\ref{omega0}).
We now scan the frequencies
$\omega_{n;i_1,\ldots,i_n;\alpha^{(n)}}$
defined in (\ref{omdef})
and collect the pairwise different ones into a set
$\Omega=\{\omega_0,\ldots,\omega_{M}\}$, where
$M+1=|\Omega|$ is the number of pairwise different frequencies
and $\omega_{M}<\omega_{M-1}<\ldots <\omega_0$.
Since
the derivation of (\ref{real3}) involved only
factoring nonzero terms out of the left-hand side of
(\ref{specdet}), the zeros of (\ref{real3}) and the
zeros of (\ref{specdet}) are identical.
Taking the real part of the real quantity (\ref{real3})
shows that
(\ref{specdet}) can be written in the form
\begin{equation}
\cos(\omega_0 k-\pi\gamma_0)-\tilde\Phi(k)=0,
\label{specequ}
\end{equation}
where
\begin{equation}
\tilde\Phi(k) =
\sum_{i=1}^{M} a_i\cos(\omega_i k-\pi\gamma_i)
\label{Phi}
\end{equation}
and $a_i, \gamma_i$, $i=0,\ldots ,M$ are real constants.
In general it is difficult to obtain
an explicit analytical result for
the zeros of (\ref{specequ}). There is, however,
a subset of quantum graphs defined in the following
section, that allows us to compute an explicit analytical
solution of (\ref{specequ}).

\section{Regular Quantum Graphs}
A subset of quantum graphs are regular quantum graphs.
They fulfil the condition
\begin{equation}
\alpha=\sum_i\mid a_i|<1,
\label{regul}
\end{equation}
where the constants $a_i$ are the coefficients of the
trigonometric
functions in
(\ref{Phi}).
Although regular quantum graphs are a restricted sub-set
of all quantum graphs, they are still quantum chaotic
with positive topological entropy \cite{PRL,PRE2,JETPL,JETP}.
Because of (\ref{regul}) we have
$|\tilde\Phi(k)|<1$ for all $k$, and the zeros $k_n$ of
(\ref{specequ}) are given by:
\begin{equation}
k_{n}=\bar k_n + \tilde k_n,\ \ \ n=1,2,\ldots,
\label{zero1}
\end{equation}
where
\begin{equation}
\bar k_n=
{\pi\over \omega_0}\left[n+\mu+{1\over 2}+\gamma_0\right]
\label{kbar}
\end{equation}
and
\begin{equation}
\tilde k_n= {(-1)^{n+\mu}\over \omega_0}\left[
\arccos[\tilde\Phi(k_n)]-{\pi\over 2}\right]
\label{ktilde}
\end{equation}
may be interpreted as the average and fluctuating parts of
the zeros of (\ref{specequ}), respectively.
Since (\ref{specequ}) is the spectral equation of a physics
problem,
we only need to study the
positive solutions of (\ref{specequ}). Therefore we introduced
the constant $\mu\in{\rm{\bf Z}}$ in (\ref{kbar})
which allows us to adjust the counting scheme of zeros
in such a way that
$k_1$ is
the first nonnegative solution of
(\ref{specequ}). This is merely a matter of
convenience and certainly
not a restriction of generality.
Because of (\ref{regul}),
the boundedness of the trigonometric functions
in (\ref{Phi}) and the properties of the $\arccos$ function,
the fluctuating part of the zeros is bounded.
We have
\begin{equation}
|\tilde k_n|\leq \tilde k_{\rm max}:={1\over\omega_0}\,
\left[{\pi\over 2}-\arccos(\alpha)\right]<{\pi\over 2\omega_0}.
\label{bound}
\end{equation}
Therefore,
roots of (\ref{specequ}) can only be found
in the intervals
$R_n=[\bar k_n-\tilde k_{\rm max},\bar k_n+\tilde k_{\rm max}]$.

We define
\begin{equation}
\hat{k}_{n}=\frac{\pi }
{\omega_0}(n+\mu+1+\gamma_0), \ \ n=1,2,\ldots
\label{k}
\end{equation}
and note in passing
that
\begin{equation}
\bar k_n=(\hat k_{n-1}+\hat k_n)/2.
\label{hbkn}
\end{equation}
We also define
the open interval
$I_n=(\hat k_{n-1},\hat k_n)$ and its
closure
$\bar I_n=[\hat k_{n-1},\hat k_n]$.
For $\alpha<1$ we have $R_n\subset I_n$.
For $\alpha\rightarrow 1$ the root intervals
$R_n$ grow in size towards $I_n$, but for any
$\alpha<1$ the end points $\hat k_{n-1}$ and
$\hat k_n$ of $\bar I_n$ are not roots of
(\ref{specequ}). The boundedness of
$\tilde k_n$ also implies the existence of two root-free
intervals in $I_n$. They are given by
$F_n^{(-)}=[\hat k_{n-1},\bar k_n-\tilde k_{\rm max}]$ and
$F_n^{(+)}=[\bar k_n+\tilde k_{\rm max},\hat k_n]$.
Thus, roots cannot be found in the union of these two
intervals, the root-free zone
$F_n=F_n^{(-)}\cup F_n^{(+)}\subset \bar I_n$.
We also have $\bar I_n=F_n\cup R_n$.
For an illustration of the various intervals defined above, and
their relation to each other, see Fig.~2.
The intervals $I_n$ together with their limiting points
$\hat k_n$
provide a natural organization of
the $k$ axis into a periodic structure of root cells.

We now define $x=\omega_0k-\pi\gamma_0$, which transforms
(\ref{specequ}) into
\begin{equation}
\cos(x)-\Phi(x)=0,\ \ \ \Phi(x)=\sum_{i=1}^M
a_i\cos(\rho_i x-\pi\nu_i),
\label{specequ'}
\end{equation}
where $\rho_i=\omega_i/\omega_0$ and $\nu_i=
\gamma_i-\rho_i\gamma_0$.
Since, as discussed in Sec.~II, $\omega_0$
is the largest frequency in (\ref{specequ}),
we have $\rho_i<1$, $i=1,\ldots,M$, and
theorem T2 (Appendix A) is applicable.
It states that there is exactly one zero $x_n$
of (\ref{specequ'})
in every open interval $(n\pi,(n+1)\pi)$, $n\in{\rm{\bf Z}}$.
Consulting Fig.~3 this fact is
intuitively clear since the $\cos$ function in
(\ref{specequ'}) is ``fast'', and $\Phi(x)$, containing
only frequencies smaller than 1, is a ``slow'' function.
Thus, as illustrated in Fig.~3,
and proved rigorously by T2 (Appendix A),
there is one and only one intersection between the fast
$\cos$ function and the slow $\Phi$ function in every
$x$ interval of length $\pi$.
Transformed back to the variable $k$ this implies that
there is exactly
one zero $k_n$ of (\ref{specequ})
in every interval $I_n$.
Since this zero cannot be found in the root-free zone
$F_n$, it has to be located in $R_n$. Thus there is
exactly one root $k_n$ of (\ref{specequ}) in
every root-interval $R_n$.
This fact is the key for obtaining
explicit analytical solutions of (\ref{specequ}) as discussed
in the following section.

\section{Periodic orbit expansions for individual spectral points}
For the zeros of (\ref{specequ})
we define the spectral staircase
\begin{equation}
N(k)=\sum_{i=1}^{\infty}\, \theta(k-k_i),
\label{stair}
\end{equation}
where
\begin{equation}
\theta(x)=\cases{0, &for $x<0$, \cr
                1/2, &for $x=0$, \cr
                 1, &for $x>0$, $x\in {\rm{\bf R}}$\cr}
\label{Heavy}
\end{equation}
is Heavyside's $\theta$ function.
Based on the scattering quantization
approach
it was shown elsewhere\cite{QGT1} that
\begin{equation}
N(k)=\bar N(k) + {1\over \pi} {\rm Im}\, {\rm Tr}\,
\sum_{l=1}^{\infty}\, {1\over l}\, S^{l}(k),
\label{N1}
\end{equation}
where
\begin{equation}
\bar N(k)= {\omega_0 k \over \pi}\ -\ (\mu+1+\gamma_0),
\label{Nbar}
\end{equation}
and $S(k)$ is the unitary scattering matrix (\ref{decomp1})
of the quantum graph.
Since, according to our assumptions, $S$ is a finite,
unitary matrix,
existence and convergence of (\ref{N1}) is guaranteed
according to L17, L18 and L19 (Appendix B).
Therefore, $N(k)$ is well-defined for all $k$.
Since $S(k)$ can easily be
constructed for any given quantum
graph\cite{JETP,QGT1},
(\ref{N1}) provides
an explicit formula for the staircase function
(\ref{stair}).
Combined with the spectral
properties of regular quantum graphs discussed in Sec.~III,
this expression now
enables us to explicitly compute the zeros of (\ref{specequ}).

In Sec.~III we proved that exactly one zero $k_n$
of (\ref{specequ}) is located in
$I_n=(\hat k_{n-1},\hat k_n)$.
Integrating $N(k)$
from $\hat k_{n-1}$ to $\hat k_n$ and taking into
account that $N(k)$ jumps by one unit at
$k=k_n$ (see illustration in Fig.~4), we obtain
\begin{equation}
\int_{\hat k_{n-1}}^{\hat k_n}\, N(k)\, dk =
N(\hat k_{n-1})[k_n-\hat k_{n-1}]+N(\hat k_n)[\hat k_n-k_n].
\label{Nint}
\end{equation}
Solving for $k_n$ and using
$N(\hat k_{n-1})=n-1$
and
$N(\hat k_n)=n$ (see Fig.~4),
we obtain
\begin{equation}
k_n={\pi\over\omega_0}\, (2n+\mu+\gamma_0)\ -\
\int_{\hat k_{n-1}}^{\hat k_n}N(k)dk.
\label{explic}
\end{equation}
Since we know $N(k)$ explicitly, (\ref{explic}) allows us to
compute every zero of (\ref{specequ}) explicitly and
individually for any choice of $n$.
The representation (\ref{explic}) requires no further
proof since, as mentioned above, $N(k)$ is well-defined
everywhere, and is Riemann-integrable over any finite
interval of $k$.

Another useful representation of $k_n$ is obtained by substituting
(\ref{N1}) with (\ref{Nbar})
into (\ref{explic}) and using (\ref{kbar}):
\begin{equation}
k_n=\bar k_n
\ -\ {1\over \pi}\, {\rm Im}\, {\rm Tr}\,
\int_{\hat k_{n-1}}^{\hat k_n}\,
\sum_{l=1}^{\infty}\, {1\over l}\, S^l(k)\, dk.
\label{explic'}
\end{equation}
According to theorem T3 (Appendix A)
presented in Sec.~V,
it is possible
to interchange integration and summation in
(\ref{explic'}) and we arrive at
\begin{equation}
k_n=\bar k_n\
\ -\ {1\over \pi}\, {\rm Im}\, {\rm Tr}\,
\sum_{l=1}^{\infty}\, {1\over l}\,
\int_{\hat k_{n-1}}^{\hat k_n} \, S^l(k)\, dk.
\label{explic''}
\end{equation}
In many cases the
integral over $S^l(k)$ can be performed explicitly, which yields
explicit representations for $k_n$.

 Finally we discuss explicit representations of $k_n$ in terms
of periodic orbits. Based on the product form (\ref{decomp1})
of the $S$ matrix and the explicit representation (\ref{DD})
of the matrix elements of $D$,
the trace
of $S(k)^l$
is of the form
\begin{equation}
{\rm Tr}\, S(k)^l = \sum_{j_1\ldots j_l}
D_{j_1,j_1}U_{j_1,j_2}D_{j_2,j_2}U_{j_2,j_3}\,
\ldots\, D_{j_l,j_l}U_{j_l,j_1}=\sum_{m\in P[l]}\, A_m[l] \,
\exp\left\{iL_m^{(0)}[l] k\right\},
\label{PO1}
\end{equation}
where $P[l]$
is the index set of all possible
periodic orbits
of length $l$ of
the graph, $A_m[l]\in {\rm {\bf C}}$
is the weight of orbit number $m$ of length $l$,
computable from
the matrix elements of $U$,
and $L_m^{(0)}[l]$ is the reduced action
of periodic orbit number $m$ of length $l$.
Using this result
together with (\ref{hbkn})
we obtain the explicit periodic orbit
formula for the spectrum in the form
\begin{equation}
k_n=\bar k_n\ -\
{2\over \pi}\, {\rm Im}\, \sum_{l=1}^{\infty}\,
{1\over l}\, \sum_{m\in P[l]}\, A_m[l]\,
{e^{iL_m^{(0)}[l]\bar k_n}\over L_m^{(0)}[l]}\,
\sin\left[{\pi\over 2\omega_0}\, L_m^{(0)}[l]\right].
\label{explic'''}
\end{equation}
Since the derivation of (\ref{explic'''}) involves only
a resummation of ${\rm Tr}\, S^l$ (which
involves only a finite number
of terms), the convergence properties of (\ref{explic''}) are
unaffected, and (\ref{explic'''}) converges.

Reviewing our logic that took us from (\ref{explic}) to
(\ref{explic'''}) it is important to stress that
(\ref{explic'''}) converges to the correct
result for $k_n$. This is so because starting from
(\ref{explic}), which we proved to be exact, we arrive
at (\ref{explic'''}) performing only allowed equivalence
transformations (as mentioned already, the
step from (\ref{explic'}) to
(\ref{explic''}) is proved in Sec.~V).
This is an important result. It means that even though
(\ref{explic'''}) may be a series that converges only
conditionally (in Sec.~VI we prove that this is
indeed the case for at least one quantum graph),
it still converges to the correct result,
provided the series is summed exactly as specified in
(\ref{explic'''}). The summation scheme specified in
(\ref{explic'''}) means that periodic orbits have to be
summed according to their symbolic lengths
(see, e.g., \cite{Devaney,Ott,BR,Gutz,JMP}) and
not, e.g., according
to their action lengths.
If this proviso is properly taken into account,
(\ref{explic'''}) is an explicit, convergent periodic orbit
representation for $k_n$ that converges to the exact
value of $k_n$.

It is possible to re-write (\ref{explic'''}) into
the more familiar form of summation over prime periodic orbits
and their repetitions.
Any periodic orbit $m$ of length $l$
in (\ref{explic'''}) consists of
an irreducible, prime periodic orbit $m_{\cal P}$ of
length $l_{\cal P}$
which is repeated $\nu$ times, such that
\begin{equation}
l=\nu l_{\cal P}.
\label{ppo1}
\end{equation}
Of course
$\nu$ may be equal to 1 if orbit number $m$ is already
a prime periodic orbit. Let us now focus on the
amplitude $A_m[l]$ in (\ref{explic''}). If we denote by
$A_{m_{\cal P}}$ the amplitude of the prime periodic orbit, then
\begin{equation}
A_m[l]=l_{\cal P}\, A_{m_{\cal P}}^{\nu}.
\label{ppo2}
\end{equation}
This is so, because the prime periodic
orbit $m_{\cal P}$ is repeated $\nu$ times, which by itself results
in the amplitude $A_{m_{\cal P}}^{\nu}$.
The factor $l_{\cal P}$ is explained in the following way:
because of the trace
in (\ref{explic''}), every vertex visited by the prime periodic
orbit $m_{\cal P}$
contributes an amplitude
$A_{m_{\cal P}}^{\nu}$ to the total amplitude
$A_m[l]$. Since the prime periodic orbit
is of length $l_{\cal P}$, i.e.
it visits $l_{\cal P}$ vertices, the total contribution is
$l_{\cal P}\, A_{m_{\cal P}}^{\nu}$.
Finally, if we denote by $L_{m_{\cal P}}^{(0)}$
the reduced action of
the prime periodic orbit $m_{\cal P}$, then
\begin{equation}
L_m^{(0)}[l] = \nu\, L_{m_{\cal P}}^{(0)}.
\label{ppo3}
\end{equation}
Collecting (\ref{ppo1}) -- (\ref{ppo3}) and inserting it into
(\ref{explic'''}) yields
\begin{equation}
k_n=\bar k_n\ -\
{2\over \pi}\, {\rm Im}\, \sum_{m_{\cal P}}\,
{1\over L_{m_{\cal P}}^{(0)}}\,
\sum_{\nu=1}^{\infty}\, {1\over \nu^2}\,
A_{m_{\cal P}}^{\nu}\,
e^{i\nu L_{m_{\cal P}}^{(0)}\bar k_n}\,
\sin\left[{\nu\pi\over 2\omega_0}\,
L_{m_{\cal P}}^{(0)}\right],
\label{explic'`}
\end{equation}
where the summation is over all prime periodic orbits
$m_{{\cal P}}$
of the graph and all their repetitions $\nu$. It is important
to note here that the summation in (\ref{explic'`}) still
has to be performed according to the symbolic lengths
$l=\nu l_{\cal P}$ of the orbits.

In conclusion we note that our
methods are generalizable to obtain
any differentiable function $f(k_n)$ directly
and explicitly.
Instead of integrating over $N(k)$ alone in (\ref{Nint})
we integrate over $f'(k)N(k)$ and obtain
\begin{equation}
f(k_n)=nf(\hat k_n)-(n-1)f(\hat k_{n-1})-
\int_{\hat k_{n-1}}^{\hat k_n}\, f'(k)\, N(k)\, dk.
\label{Nfint}
\end{equation}
Following the same logic that led to (\ref{explic'''}),
we obtain
\begin{equation}
f(k_n)=nf(\hat k_n)-(n-1)f(\hat k_{n-1})-
{2\over \pi}\, {\rm Im}\, \sum_{l=1}^{\infty}\,
{1\over l}\, \sum_{m\in P[l]}\, A_m[l]\,
G_n(L_m^{(0)}[l]),
\label{explicf}
\end{equation}
where
\begin{equation}
G_n(x)\ = \
\int_{\hat k_{n-1}}^{\hat k_n}\, f'(k)\,
e^{ixk}\, dk.
\end{equation}
This amounts to a resummation
since one can also
obtain the series for
$k_n$ first, and then form $f(k_n)$.

\section{Interchange of Integration and Summation}
One of the key points for the existence of
the explicit formula (\ref{explic'''})
is the
possibility to interchange integration and summation
according to
\begin{equation}
 \int_a^b\left(\sum_{n=1}^{\infty}\,
 {e^{in\sigma(x)}\over n}\right)\, dx =
\sum_{n=1}^{\infty}\left(
 \int_a^b\, {e^{in\sigma(x)}\over n}\, dx\right),
\label{inter}
\end{equation}
where $\sigma(x)$ is continuous and has a
continuous first derivative.
We will prove (\ref{inter}) in two steps.

\underbar{\it Step 1:}
According to a well-known theorem (see, e.g.,
Ref. \cite{RC}, volume I, p. 394)
summation and integration can be interchanged if the
convergence of the sum
is uniform.
Consider an interval $[\sigma_1,\sigma_2]$
that does not contain a point $\sigma_0=0$ mod $2\pi$.
Define
\begin{equation}
{\cal S}(\sigma)=\sum_{n=1}^{\infty}\, {e^{in\sigma}\over n}.
\label{calS}
\end{equation}
Let
\begin{equation}
f(\sigma)=
       {1\over 2}\ln{1\over 2[1-\cos(\sigma)]}
+i{\pi-\sigma\,{\rm mod}\, 2\pi\over 2}.
\end{equation}
Then, according to formulas F2
and F3 (Appendix B),
${\cal S}(\sigma)=f(\sigma)$ in $[\sigma_1,\sigma_2]$.
In other words, ${\cal S}(\sigma)$ is the Fourier
series representation
of $f(\sigma)$. According to another
well-known
theorem (see, e.g., Ref. \cite{CH},
volume I, pp. 70--71) the
Fourier series of a piecewise continuous function converges
uniformly in every closed interval in which the function
is continuous.
Since $f(\sigma)$ is continuous and smooth in
$[\sigma_1,\sigma_2]$, ${\cal S}(\sigma)$ converges
uniformly in $[\sigma_1,\sigma_2]$. This means that
summation and integration can be interchanged in any
closed interval $[x_1,x_2]$ for which
$\sigma(x)\neq 0$ mod $2\pi$ $\forall x\in [x_1,x_2]$.

\underbar{\it Step 2:}
Now let there be a single point $x^*\in
(x_1,x_2)$ with $\sigma(x^*)=0$ mod $2\pi$. Then, for any
$\epsilon_1,\epsilon_2>0$ with $x^*-\epsilon_1
\geq x_1$, $x^*+\epsilon_2\leq x_2$, ${\cal S}(\sigma(x))$
is uniformly convergent in
$[x_1,x^*-\epsilon_1]$ and $[x^*+\epsilon_2,x_2]$
and integration and summation can
be interchanged when integrating over these two intervals.
Consequently,
$$
\int_{x_1}^{x_2}\,
\left(\sum_{n=1}^{\infty}\, {e^{in\sigma(x)}\over n}
\right)\, dx\ =\ $$
$$
\int_{x_1}^{x^*-\epsilon_1}\,
\left(\sum_{n=1}^{\infty}\, {e^{in\sigma(x)}\over n}
\right)\, dx\ +\
\int_{x^*-\epsilon_1}^{x^*+\epsilon_2}\,
\left(\sum_{n=1}^{\infty}\, {e^{in\sigma(x)}\over n}
\right)\, dx\ +\
\int_{x^*+\epsilon_2}^{x_2}\,
\left(\sum_{n=1}^{\infty}\, {e^{in\sigma(x)}\over n}
\right)\, dx\ =$$
\begin{equation}
\sum_{n=1}^{\infty}\left(
 \int_{x_1}^{x^*-\epsilon_1}\,
 {e^{in\sigma(x)}\over n}\, dx\right)\ +\
\sum_{n=1}^{\infty}\left(
 \int_{x^*+\epsilon_2}^{x_2}\,
 {e^{in\sigma(x)}\over n}\, dx\right)\ +\
\int_{x^*-\epsilon_1}^{x^*+\epsilon_2}\,
\left(\sum_{n=1}^{\infty}\, {e^{in\sigma(x)}\over n}
\right)\, dx.
\end{equation}
Since $\sum_{n=1}^{\infty}\int_{x_1}^{x^*-\epsilon_1}
\exp[in\sigma(x)]/n\, dx$ and
$\sum_{n=1}^{\infty}\int_{x^*+\epsilon_2}^{x_2}
\exp[in\sigma(x)]/n\, dx$ are both uniformly convergent,
we have with L11 (Appendix B):
$$
\int_{x_1}^{x_2}\,
\left(\sum_{n=1}^{\infty}\, {e^{in\sigma(x)}\over n}
\right)\, dx\ =\
$$
\begin{equation}
\sum_{n=1}^{\infty}
\left[
\int_{x_1}^{x^*-\epsilon_1}\, {e^{in\sigma(x)}\over n}\, dx\ +\
\int_{x^*+\epsilon_2}^{x_2}\,
{e^{in\sigma(x)}\over n}\, dx \right]
+ \int_{x^*-\epsilon_1}^{x^*+\epsilon_2}\,
\left(\sum_{n=1}^{\infty}\, {e^{in\sigma(x)}\over n}
\right)\, dx.
\label{fpf}
\end{equation}
Since $\exp[in\sigma(x)]/n$ is a non-singular, smooth
function at $x=x^*$, there is no problem with
taking $\epsilon_1,\epsilon_2\rightarrow 0$ for the
first two integrals on the right-hand side of
(\ref{fpf}).
Therefore, integration and summation on the left-hand side
of (\ref{fpf}) can be interchanged if
\begin{equation}
\lim_{\epsilon_1,\epsilon_2\rightarrow 0}\,
 \int_{x^*-\epsilon_1}^{x^*+\epsilon_2}\,
\left(\sum_{n=1}^{\infty}\, {e^{in\sigma(x)}\over n}
\right)\, dx\ =\ 0\ =\
\lim_{\epsilon_1,\epsilon_2\rightarrow 0}\,
 \sum_{n=1}^{\infty} \left( \int_{x^*-\epsilon_1}^{x^*+\epsilon_2}\,
{e^{in\sigma(x)}\over n}\, dx
\right).
\end{equation}
This is guaranteed according to T3 (Appendix A).
Assuming that $\sigma(x)$ has only a finite number $N$
of zeros mod $2\pi$ in $(a,b)$, we can break $(a,b)$
into $N$ sub-intervals containing a single zero only,
in which the interchange of summation and integration
is allowed. This proves (\ref{inter}).

Returning to the crucial step from (\ref{explic'}) to
(\ref{explic''}) we have to show that
\begin{equation}
\int_{k_1}^{k_2}\,
\sum_{n=1}^{\infty}\,
{1\over n}\, S^n(k)\, dk =
\sum_{n=1}^{\infty}\, {1\over n}\,
\int_{k_1}^{k_2}\, S^n(k)\, dk.
\end{equation}
Since $S(k)$ is unitary, it is diagonalizable, i.e.
there exists a matrix $W(k)$ such that
\begin{equation}
S(k)=W(k)\, {\rm diag}\left(e^{i\sigma_1(k)},\ldots,
e^{i\sigma_{2N_B}(k)}\right)\, W^{\dagger}(k),
\end{equation}
where $\sigma_1(k),\ldots,\sigma_{2N_B}(k)$ are the
$2N_B$ eigenphases of $S(k)$. Because of the
structure (\ref{decomp1})
of the $S$ matrix in conjunction with
the smoothly varying phases (\ref{DD}),
the eigenphases of the $S$ matrix have only a
finite number of zeros mod $2\pi$ in any finite
interval of $k$. This is important for later
use of (\ref{inter}) which was only proved for
this case.

We now make essential use of our focus on
finite quantum graphs, which entails a finite-dimensional
$S$ matrix, and therefore a finite-dimensional
matrix $W$.
In this case matrix multiplication with $W$ leads only
to finite sums. Since for finite sums
integration and
summation is always interchangeable we have
$$
\int_{k_1}^{k_2}\,
\sum_{n=1}^{\infty}\,
{1\over n}\, S^n(k)\, dk =
\int_{k_1}^{k_2}\, W(k)
\sum_{n=1}^{\infty}\,
{\rm diag}\left({e^{in\sigma_1(k)}\over n},\ldots,
{e^{in\sigma_{2N_B}(k)}\over n}\right)\,
W^{\dagger}(k)\, dk \
{\buildrel (\ref{inter}) \over =}
$$
\begin{equation}
\sum_{n=1}^{\infty}\,
\int_{k_1}^{k_2}\, W(k)
{\rm diag}\left({e^{in\sigma_1(k)}\over n},\ldots,
{e^{in\sigma_{2N_B}(k)}\over n}\right)\,
W^{\dagger}(k)\, dk \ =
\sum_{n=1}^{\infty}\, {1\over n}\,
\int_{k_1}^{k_2}\, S^n(k)\, dk.
\end{equation}
This equation justifies the step from
(\ref{explic'}) to
(\ref{explic''}), which proves
the validity of (\ref{explic'''}) and (\ref{explic'`}).

\section{Convergence properties of the periodic orbit series}
In this section we
prove rigorously that (\ref{explic'`}) contains
conditionally convergent as well as absolutely convergent
cases. We accomplish this by investigating
the convergence properties of (\ref{explic'`})
in the case of the
dressed three-vertex linear graph shown in Fig.~5.
The potential on the bond $B_{12}$ is zero; the
potential on the bond $B_{23}$ is a scaling potential
explicitly given by
\begin{equation}
U_{23}= \lambda E,
\label{pot}
\end{equation}
where $E$ is the energy of the quantum particle and
$\lambda$ is a real constant with $0<\lambda<1$.
The quantum graph shown in Fig.~5
was studied in detail in
\cite{PRL,PRE1,PRE2,JETPL,JETP,JMP}.
Denoting by $a$ the geometric length of the bond
$B_{12}$ and by $b$ the geometric length of the bond
$B_{23}$, its spectral
equation is given by
\cite{PRL,PRE1,PRE2,JETPL,JETP,JMP}
\begin{equation}
\sin(\omega_0 k)=r\sin(\omega_1 k),
\label{3specequ}
\end{equation}
where
\begin{equation}
\omega_0=a+\beta b,\ \ \ \omega_1=
a-\beta b,\ \ \ r={1-\beta
\over 1+\beta},\ \ \ \beta=\sqrt{1-\lambda}.
\label{omegas}
\end{equation}
With
\begin{equation}
 \gamma_0=1/2,\ \ \ a_1=r,\ \ \ \gamma_1=1/2,
\label{params}
\end{equation}
the spectral equation (\ref{3specequ}) is precisely
of the form (\ref{specequ}).
Since according to
(\ref{omegas}) $a_1=r<1$, the regularity condition
(\ref{regul}) is fulfilled and
(\ref{3specequ}) is the
spectral equation of a regular quantum graph.
This means that we can apply (\ref{explic'`}) for
the computation of the solutions of (\ref{3specequ}).
In order to do so, we need a scheme for enumerating
the periodic orbits of the three-vertex linear graph.
It was shown in \cite{JMP} that a one-to-one
correspondence exists between the periodic orbits of
the three-vertex linear graph and the set of
binary P\'olya necklaces \cite{JMP,Comb1,Comb2}.
A binary necklace is a string of two symbols arranged
in a circle such that two necklaces are different
if (a) they are of different lengths or (b)
they are of the same length but cannot be made to
coincide even after applying cyclic shifts of the
symbols of one of the necklaces. For the graph
of Fig.~5 it is convenient to introduce the two
symbols ${\cal L}$ and ${\cal R}$, which can be
interpreted
physically as the reflection of a graph particle from
the left ($V_1$)
or the right ($V_3$)
dead-end vertices, respectively.
Since strings of symbols are frequently referred to as
words, we adopt the symbol $w$ to denote a particular
necklace. For a given necklace $w$ it is convenient
to define the following functions\cite{JMP}:
$n_{\cal R}(w)$, which counts
the number of ${\cal R}$s in $w$, $n_{\cal L}(w)$,
which counts the number of ${\cal L}$s in $w$, the
pair function $\alpha(w)$, which counts all occurrences
of ${\cal R}$-pairs or ${\cal L}$-pairs in $w$
and the function
$\beta(w)$, which counts all occurrences of ${\cal LR}$ or
${\cal RL}$ symbol combinations in $w$. We also define the function
$\ell(w)=n_{\cal L}(w)+n_{\cal R}(w)$,
which returns the
total binary length of the word $w$, and the phase
function $\chi(w)$, defined as the sum of $\ell(w)$ and
the number of ${\cal R}$-pairs in $w$.
We note the identity
\begin{equation}
\alpha(w)+\beta(w)=\ell(w).
\label{ident}
\end{equation}
In evaluating the functions defined above, we have to
be very careful to take note of the cyclic nature of
binary necklaces. Therefore, for example,
$\alpha({\cal R})=1$, $\beta({\cal LR})=2$,
$\alpha({\cal LLRRL})=3$ and $\beta({\cal LLRRL})=2$,
which also checks (\ref{ident}).
In addition we
define the set $W(l)$ of all binary
necklaces of length $l$.

Let us look at $W(2)$. This set contains three necklaces,
${\cal LL}$, ${\cal LR}={\cal RL}$ (cyclic rotation of
symbols) and ${\cal RR}$. The necklace ${\cal LL}$ is
not a primitive necklace, since it consists of a repetition
of the primitive
symbol ${\cal L}$. The same holds for the
necklace ${\cal RR}$, which is a repetition of the
primitive symbol ${\cal R}$. The necklace ${\cal LR}$
is primitive, since it cannot be written as a
repetition of a shorter string of symbols.
This motivates the definition of the set
$W_{\cal P}$ of all primitive binary necklaces
and the set $W_{\cal P}(l)$ containing
all primitive binary necklaces of length $l$.

An important question arises: How many primitive
necklaces $N_{\cal P}(l)$
are there in $W(l)$? In other words, how many members
are there in $W_{\cal P}(l)\subset W(l)$?
The following formula gives the answer\cite{Comb2}:
\begin{equation}
N_{\cal P}(l) = {1\over l}\, \sum_{m|l}\,
\phi(m)\, 2^{l/m},
\label{nneck}
\end{equation}
where the symbol ``$m|l$'' denotes ``$m$ is a divisor
of $l$'', and
$\phi(m)$ is Euler's totient function defined
as the
number of positive integers smaller than $m$ and relatively
prime to $m$ with $\phi(1)=1$ as a useful convention.
It is given explicitly by \cite{AS}
\begin{equation}
\phi(1)=1,\ \ \ \phi(n)=n\, \prod_{p|n}\,
\left(1-{1\over p}\right),\ \ n\geq 2,
\label{Eultot}
\end{equation}
where $p$ is a prime number.
Thus the first four totients are given by $\phi(1)=1$,
$\phi(2)=1$, $\phi(3)=2$ and $\phi(4)=2$.

A special case of (\ref{nneck}) is the case in which
$l$ is a prime number. In this case we have explicitly
\begin{equation}
N_{\cal P}(p) = {1\over p}\, \left(2^p-2\right),\ \ \
p\ {\rm prime}.
\label{nneck'}
\end{equation}
This is immediately obvious from the following
combinatorial
argument. By virtue of
$p$ being a prime number a necklace of length $p$ cannot
contain an integer repetition of shorter substrings, except
for strings of length 1 or length $p$. Length $p$ is trivial.
It corresponds to the word itself. Length 1 leads to the
two cases ${\cal RRRRR...R}$ and ${\cal LLLLL...L}$, where
the symbols ${\cal R}$ and ${\cal L}$, respectively, are
repeated $p$ times. So, except for these two special
necklaces, any necklace of prime length $p$ is automatically
primitive. Thus there are
\begin{equation}
{1\over p}\, \left(\matrix{p\cr \nu\cr}\right)
\label{nlct}
\end{equation}
different necklaces with $\nu$ symbols ${\cal L}$ and
$p-\nu$ symbols ${\cal R}$, where the factor $1/p$ takes care
of avoiding
double counting of cyclically equivalent necklaces.
In total, therefore, we have
\begin{equation}
N_{\cal P}(p)={1\over p}\, \sum_{\nu=1}^{p-1}\,
\left(\matrix{p\cr \nu\cr}\right)={1\over p}\,
\left(2^p-2\right)
\label{nneck''}
\end{equation}
primitive necklaces of length $p$, in agreement with
(\ref{nneck'}). The sum in (\ref{nneck''}) ranges
from 1 to $p-1$ since $\nu=0$ would correspond to
the composite, non-primitive necklace ${\cal RRRR...R}$ and
$\nu=p$ would correspond to the composite, non-primitive necklace
${\cal LLLLL...L}$.

We are now ready to apply (\ref{explic'`}) to
the three-vertex linear graph. In ``necklace notation''
it is given by
\cite{PRL,PRE1,PRE2,JETPL,JETP,JMP}
\begin{equation}
k_n=\bar k_n\ -\
{2\over \pi}\, \sum_{l=1}^{\infty}
\sum_{\nu=1}^{\infty}\,
\sum_{w\in W_{\cal P}:\nu w\in W(l)}\,
{A_w^{\nu}\over \nu^2\, L_w^{(0)}}\,
\sin\left[\nu L_w^{(0)}\bar k_n\right]\,
\sin\left[{\nu\pi\over 2\omega_0}\, L_w^{(0)}\right],
\label{explic(1)}
\end{equation}
where $L_w^{(0)}$ is the reduced action of the primitive
necklace $w$, given by \cite{JMP}
\begin{equation}
L_w^{(0)}=2[n_{\cal L}(w)a + n_{\cal R}(w)\beta b]
\label{redact}
\end{equation}
and the amplitude $A_w$ of the primitive necklace $w$
is given by \cite{JMP}
\begin{equation}
A_w=
(-1)^{\chi(w)}\, r^{\alpha(w)}\, (1-r^2)^{\beta(w)/2},
\label{ampl}
\end{equation}
where $r$ and $\beta$ are defined in (\ref{omegas}).
The notation
$\nu w$ refers to a necklace of binary length $\nu \ell(w)$
that consists of
$\nu$ concatenated substrings $w$.
Note that the summations in (\ref{explic(1)}) are
ordered in such a way that for fixed $l$ we sum
over all possible primitive words $w$ and their repetitions
$\nu$ such that the total length of the resulting
binary necklace amounts to $l$, and only then do we
sum over the binary length $l$ of the necklaces.
This summation scheme, explicitly specified in (\ref{explic(1)}),
complies completely with the
summation scheme defined in Sec.~IV.
Since we proved in Sec.~IV that (\ref{explic'`}) converges,
provided we adhere to the correct summation scheme, so
does
(\ref{explic(1)}).

A numerical example of the computation
of the spectrum of (\ref{3specequ}) via
(\ref{explic(1)})
was presented
in \cite{PRL} where we chose
$a=0.3$, $b=0.7$ and $\lambda=1/2$. For $n=1$, 10, 100
we computed the exact
roots of (\ref{3specequ}) numerically by using a simple
numerical root-finding algorithm. We
obtained
$k_1^{({\rm exact})}\approx$ 4.107149,
$k_{10}^{({\rm exact})}\approx$ 39.305209
and $k_{100}^{({\rm exact})}\approx$ 394.964713.
Next we computed these roots using the explicit
formula (\ref{explic(1)}).
Including all binary necklaces up to $l=20$,
which amounts to including a total of approximately
$10^5$ primitive periodic necklaces,
we obtained
$k_1^{({\rm p.o.})}\approx$ 4.105130,
$k_{10}^{({\rm p.o.})}\approx$ 39.305212
and $k_{100}^{({\rm p.o.})}\approx$ 394.964555.
Given the fact that in Sec.~IV
we proved exactness and convergence
of (\ref{explic'`}) ((\ref{explic(1)}), respectively),
the good agreement
between $k_n^{({\rm exact})}$ and
$k_n^{({\rm p.o.})}$, $n=1,10,100$,
is not surprising. Nevertheless we found it
important to present this simple example here,
since it illustrates the abstract procedures
and results obtained in Sec.~IV,
checks our algebra and
instills confidence in our methods.

We now investigate the convergence properties of
(\ref{explic(1)}) for
two special cases of dressed linear
three-vertex quantum graphs (see Fig.~5)
defined by
\begin{equation}
r={1\over\sqrt{2}},\ \ \ a=m\beta b,\ \ m=1,2.
\label{fampar}
\end{equation}
In this case the reduced actions (\ref{redact})
reduce to
\begin{equation}
L_w^{(0)}=2a[n_{\cal L}(w) + n_{\cal R}(w)/m]
\label{redact'}
\end{equation}
and $\omega_0$ is given by
\begin{equation}
\omega_0=a\left(1+{1\over m}\right).
\label{om0'}
\end{equation}
Using (\ref{ident}),
the amplitudes (\ref{ampl}) are
\begin{equation}
A_w= (-1)^{\chi(w)}\, 2^{-\ell(w)/2}.
\label{ampl'}
\end{equation}
We now show that for $m=1$ the first $\sin$-term
in (\ref{explic(1)}) is always zero, and thus
(\ref{explic(1)}) converges (trivially) absolutely
in this case. For $m=1$ (\ref{redact'}) becomes
\begin{equation}
L_w^{(0)}=2a[n_{\cal L}(w) + n_{\cal R}(w)]=2a\ell(w).
\label{redact''}
\end{equation}
Also, according to (\ref{kbar}) and
(\ref{params}) $\bar k_n$
is given by
\begin{equation}
\bar k_n=
{\pi\over \omega_0}\left[n+\mu+1\right].
\label{kbar'}
\end{equation}
Thus, for $m=1$
the argument of the first $\sin$-term in
(\ref{explic(1)}) is given by
\begin{equation}
\nu L_w^{(0)}\bar k_n=\nu \ell(w) (n+\mu+1) \pi.
\end{equation}
This is an integer multiple of $\pi$, and thus
all terms in the periodic-orbit sum of
(\ref{explic(1)}) vanish identically.
Therefore we proved that there exists at least one
case in which
the periodic-orbit sum in
(\ref{explic(1)}) is (trivially) absolutely convergent.

We now prove rigorously that there exists
at least one non-trivial
case in which (\ref{explic(1)}) converges only
conditionally.
Since we already proved in Sec.~IV that
(\ref{explic(1)})
always converges,
all we have to prove is that
there exists a case in which the sum of
the absolute values of the
terms in (\ref{explic(1)}) diverges.
In order to accomplish this, let us focus on
the case $m=2$ and estimate the sum
\begin{equation}
s= \sum_{l=1}^{\infty}
\sum_{\nu=1}^{\infty}\,
\sum_{w\in W_{\cal P}:\nu w\in W(l)}\,
\left|
{1\over \nu^2\, L_w^{(0)}2^{\ell(w)/2}}\,
\sin\left[\nu L_w^{(0)}\bar k_n\right]\,
\sin\left[{\nu\pi\over 2\omega_0}\, L_w^{(0)}\right]
\right|.
\label{explic(2)}
\end{equation}
We now restrict the summation over all integers $l$
to the summation over prime numbers $p$ only.
Moreover, we discard all non-primitive necklaces
of length $p$, which is equivalent to
keeping terms
with $\nu=1$ only.
Observing that trivially
$\ell(w)=p$ for all necklaces in $W_{\cal P}(p)$,
we obtain:
\begin{equation}
s\geq \sum_p
\sum_{w\in W_{\cal P}(p)}\,
\left|
{1\over \, L_w^{(0)}2^{p/2}}\,
\sin\left[L_w^{(0)}\bar k_n\right]\,
\sin\left[{\pi\over 2\omega_0}\, L_w^{(0)}\right]
\right|,
\label{explic(3)}
\end{equation}
where the sum is over all prime numbers $p$.
For $m=2$
the reduced actions are given by
\begin{equation}
L_w^{(0)}=a[2n_{\cal L}(w) + n_{\cal R}(w)]
\label{redact'''}
\end{equation}
and
\begin{equation}
\omega_0={3a\over 2},\ \ \
\bar k_n={2\pi\over 3a}(n+\mu+1).
\end{equation}
We use these relations to evaluate the arguments
of the two $\sin$-functions
in (\ref{explic(3)}). We obtain
\begin{equation}
L_w^{(0)}\bar k_n={2\pi\over 3}
[2n_{\cal L}(w) + n_{\cal R}(w)]\, (n+\mu+1)
\label{Psi1}
\end{equation}
and
\begin{equation}
{\pi\over 2\omega_0}\, L_w^{(0)}={\pi\over 3}
[2n_{\cal L}(w) + n_{\cal R}(w)],
\label{Psi2}
\end{equation}
respectively.
We see immediately that all terms in (\ref{explic(3)})
are zero if $n+\mu+1$ is divisible by 3. This provides
additional examples of (trivially) absolutely
convergent cases of (\ref{explic(1)}).
In case $n+\mu+1$ is not divisible by 3, only those
terms contribute to (\ref{explic(3)}) for which
$2n_{\cal L}(w) + n_{\cal R}(w)$ is not divisible by 3.
Following the reasoning that led to (\ref{nneck''}),
$n_{\cal L}(w)$ ranges from 1 to $p-1$
for $w\in W_{\cal P}(p)$. Then,
$2n_{\cal L}(w) + n_{\cal R}(w)$ ranges from $p+1$
to $2p-1$ in steps of 1. Since $p+3j$ is never
divisible by 3 for $p$ prime and $j\in {\rm{\bf N}}$,
the number of primitive
necklaces $w$ of length $p$
with the property that
$2n_{\cal L}(w) + n_{\cal R}(w)$ is not divisible by 3
is at least
\begin{equation}
{1\over p}\, \left\{
\left(\matrix{p\cr 3\cr}\right)+
\left(\matrix{p\cr 6\cr}\right)+\ldots
\right\}=
{1\over 3p}\,
\left[2^p+2\cos\left({p\pi\over 3}\right)\, -\, 3\right] ,
\label{subsm}
\end{equation}
where the sum over the binomial coefficients was
evaluated with the help of formula 0.1521 in
\cite{GR}.
Therefore, with (\ref{redact'''}), (\ref{Psi1}),
(\ref{Psi2}),
(\ref{subsm}),
$|\sin(2j\pi/3)|=\sqrt{3}/2$ for all $j\in {\rm{\bf N}}$
and
$2n_{\cal L}(w) + n_{\cal R}(w)\leq 2p-1$ for
$w \in W_{\cal P}(p)$,
we obtain
\begin{equation}
s\geq {1\over 4a}\sum_p
{1\over p(2p-1)\, 2^{p/2}}\,
\left[2^p+2\cos\left({p\pi\over 3}\right)\, -\, 3\right] ,
\label{explic(4)}
\end{equation}
which obviously diverges exponentially. The physical
reason is that the quantum amplitudes, which
contribute the factor $2^{-p/2}$ in (\ref{explic(4)})
are not able to counteract the proliferation
$2^p$ of
primitive periodic orbits (primitive binary necklaces)
in (\ref{explic(4)}).
Analogous results can easily be obtained for
graphs with $m>2$ in (\ref{fampar}).

In summary we established in this section that
the convergence properties of (\ref{explic'`})
depend on the details of the quantum graph
under investigation. We proved rigorously
that both
conditionally convergent
and absolutely convergent cases
can be found.
We emphasize that the degree of convergence does not
change the fact, proved in Sec.~IV, that (\ref{explic'`})
always
converges, and always
converges to the exact spectral eigenvalues.

\section{Lagrange's inversion formula}
The periodic orbit expansions presented in Sec.~IV
are not the only way
to obtain the spectrum of regular quantum graphs
explicitly.
Lagrange's inversion formula \cite{LagInv}
offers an alternative route.
Given an implicit equation of the form
\begin{equation}
x = a + w\varphi(x)
\label{Lagr1},
\end{equation}
Lagrange's inversion formula determines a root
$x^*$ of (\ref{Lagr1}) according to
the explicit series expansion
\begin{equation}
x^*=a+\sum_{\nu=1}^{\infty}\, {w^{\nu}\over \nu !}\,
{d^{\nu-1}\over dx^{\nu-1}}\, \varphi^{\nu}(x)\Big|_a,
\label{Lagr2}
\end{equation}
provided $\varphi(x)$ is analytic in an open interval $I$
containing $x^*$ and
\begin{equation}
|w|\ <\ \left|{x-a\over \varphi(x)}\right|\ \ \forall\ x\in I.
\label{Lagr3}
\end{equation}
Since (\ref{zero1}) is of the form (\ref{Lagr1}), and
the regularity condition (\ref{regul}) ensures that
(\ref{Lagr3}) is satisfied, we can use
Lagrange's inversion formula (\ref{Lagr2}) to compute
explicit solutions of (\ref{specequ}).

In order to illustrate Lagrange's inversion
formula we will now apply it to the solution of
(\ref{3specequ}).
Defining $x=\omega_0 k$, the $n$th root of
(\ref{3specequ}) satisfies the implicit equation
\begin{equation}
x_n = \pi n + (-1)^n\,  \arcsin[r\sin(\rho x_n)],
\label{impl}
\end{equation}
where $\rho=\omega_1/\omega_0$ and $|\rho|<1$.
For the same parameter values
as specified in \cite{PRL} and already used above in
Sec.~VI we obtain
$x_1^{({\rm exact})}=\omega_0 k_1^{({\rm exact})}
\approx 3.265080$,
$x_{10}^{({\rm exact})}=\omega_0 k_{10}^{({\rm exact})}
\approx 31.246649$ and
$x_{100}^{({\rm exact})}=\omega_0 k_{100}^{({\rm exact})}
\approx 313.986973$.
We now re-compute these values using
the first two terms in the expansion (\ref{Lagr2}).
For our example they
are given by
\begin{equation}
x_n^{(2)}=\pi n + \arcsin[r\sin(\rho\pi n)]
\left\{(-1)^n + {r\rho\cos(\rho\pi n)\over
\sqrt{1-r^2\sin^2(\rho\pi n)}}\right\}.
\label{Lagexpl}
\end{equation}
We obtain $x_1^{(2)}=3.265021\ldots$,
$x_{10}^{(2)}=31.246508\ldots$ and
$x_{100}^{(2)}=313.986819\ldots$,
in very good agreement with
$x_1^{({\rm exact})}$,
$x_{10}^{({\rm exact})}$ and
$x_{100}^{({\rm exact})}$.

Although both, (\ref{explic'`}) and (\ref{Lagr2}) are
exact, and, judging from our example,
(\ref{Lagr2}) appears to converge very quickly,
the main difference between (\ref{explic'`}) and
(\ref{Lagr2}) is that no physical insight can be obtained from
(\ref{Lagr2}), whereas (\ref{explic'`}) is tightly connected
to the classical mechanics of the graph system providing,
in the spirit of Feynman's path integrals,
an intuitively clear picture of the physical processes
in terms of a superposition of amplitudes associated
with classical periodic orbits.

\section{Discussion, summary and conclusion}
There are only very few exact results in quantum chaos
theory. In particular not much is known about
the convergence properties of periodic orbit expansions.
Since quantum graphs are an important model for
quantum chaos \cite{KS},
which in fact
have already been called ``paradigms of quantum chaos''
\cite{paradigm},
it seems natural that they
provide a logical starting point for the
mathematical investigation of quantum chaos.
The regular quantum graphs defined in this paper are
important because they provide the first example
of an explicitly solvable quantum chaotic system. Moreover
regular quantum graphs allow us to prove two important
results: (a) Not all periodic orbit expansions diverge.
There exist nontrivial,
convergent, periodic orbit expansions. (b)
There exist explicit periodic orbit expansions
that converge to the exact values of individual
spectral points.

The main result of this paper
is an analytical proof of the validity and the convergence
of the explicit spectral formulas (\ref{explic'''}) and
(\ref{explic'`}), respectively. This result is novel
in two respects. (i) While periodic orbit expansions of
the spectral density and the spectral staircase of
a quantum system are basic tools of quantum chaos,
the very concept of a
periodic orbit expansion for individual spectral
eigenvalues is new. (ii) Due to the exponential
proliferation of the number of periodic orbits with
their (action) lengths, it is frequently assumed in
the quantum chaos community that periodic orbit
expansions are formal tools at best, but do not
converge. We proved in this paper that, at least as
far as regular quantum graphs are concerned, and
despite the exponential proliferation of periodic
orbits in this case \cite{PRL}, the periodic orbit
expansion (\ref{explic'`}) converges in the usual
sense of elementary analysis. This result is also new.

The main ingredient in the proof of (\ref{explic'`}) is
theorem T2 (Appendix A), i.e. an analytical proof that there
is exactly one spectral point in every root cell $I_n$.
In discussions with our colleagues we found that
while many pointed out the necessity of justifying
the interchange of
integration and summation in (\ref{inter}) (now
established in Sec.~V with T3 (Appendix A)),
many were initially puzzled by
the existence of root intervals
and the organization of the spectral points into
root cells, now
guaranteed by T2 (Appendix A). This is so because
regular quantum graphs have a positive topological
entropy \cite{Ott,Gutz,PRL} and are in this sense
quantum chaotic systems. Hence
the spectrum of
regular quantum graphs is expected to be ``wild'',
in complete contrast to the fact, proved in this paper,
that the spectrum
of regular quantum graphs can actually be organized
into regular root cells.
In this sense regular quantum graphs are closely related
to other quantum chaotic systems that also show
marked deviations from the expected universal behavior
\cite{Gutz,STOECK,Haake}.
As a specific example we mention chaotic billiards
on the hyperbolic plane generated
by arithmetic groups \cite{BGGS}.
We hope that the
pedagogical presentation of the proofs in Appendices A and B,
with their hierarchical structure and the use of only
elementary analysis concepts will help to establish
theorems T2 and
T3 (Appendix A), and their
consequence, the existence of explicit,
convergent periodic orbit expansions.
We mention that the
spectral equation
(\ref{specequ}) of a finite quantum graph is an
example of an almost periodic function \cite{HBohr}.
More information on the
analytical structure of the zeros of almost periodic
functions can be found in \cite{Zerobook}.

There are many basic quantum
mechanical problems that lead to transcendental
equations of the type (\ref{specequ}).
So far
the recommended method is to solve them graphically
or numerically (see, e.g., \cite{Schiff,Mess}).
Based on the results presented in this paper,
a third method is now available for presentation
in text books on quantum mechanics: explicit analytical
solutions.
When the regularity condition
(\ref{regul}) is satisfied, either
the Lagrangian inversion
method or the periodic orbit expansion
(\ref{explic'`}) may be employed. Since the
Lagrangian method is a purely mathematical tool
without immediate physical meaning, the periodic
orbit expansion may be preferred due to its
direct physical relevance in terms of concrete
classical physics.

Having been established with mathematical rigor
in this paper,
formula (\ref{explic'`}) may serve as
the starting point for many further investigations.
We mention one: Since according to (\ref{explic'`})
$k_n$ is known explicitly, so is the level
spacing $s_n=k_n-k_{n-1}$. This may give us an
important handle
on investigating analytically and exactly
the nearest-neighbor spacing statistics $P(s)$
\cite{Gutz,Haake}
of regular quantum graphs.
Whatever the precise properties of $P(s)$
will be, one result is clear already: due to
the existence of the root-free zones $F_n$, established
in Sec.~III, $P(s)$ is not Wignerian.
Thus, regular quantum graphs will join the growing
class of classically chaotic systems which
do not show the generic properties of typical
quantum chaotic systems.

A corollary of some significance is the following.
Since we proved that
for regular quantum graphs
there is exactly one root $k_n$ of (\ref{specequ})
in $I_n$, this
proves rigorously that
for regular quantum graphs
the number of roots of (\ref{specequ}) smaller than
$k$ grows like $\bar N(k)\sim \omega_0 k/\pi$
(Weyl's law \cite{Gutz}).

An open question is the generalization of our results
to the case of infinite quantum graphs. In case
$\sum_{i=1}^{\infty} |a_i|$ converges, it seems
straightforward to generalize the regularity condition
(\ref{regul}) to the case of infinite quantum graphs.

In summary we proved a rigorous theorem on the existence
and convergence of explicit periodic orbit expansions
of the spectral points of regular quantum graphs.
We hope that this paper will lay the foundation for
further rigorous research in quantum graph theory.

\section{Acknowledgments}
Y.D. and R.B. gratefully acknowledge financial support by NSF grant
PHY-9900730 and PHY-9984075; Y.D. by NSF grant PHY-9900746.

\section{Appendix A: Theorems}

\noindent
\underbar{\bf Theorem T1:}
Let $a_i$, $\omega_i$,
$\alpha_i\in {\rm{\bf R}}$,
$i\in I:=\{1,\ldots,N\}$, $N\in {\rm {\bf N}}$,
$\sum_{i=1}^N |a_i|<1$,
and $|\omega_i|<1$. Then:
\begin{equation}
f(x):={\left[\sum_{i=1}^N a_i\omega_i\sin(\omega_i x+\alpha_i)
\right]^2\over
1-\left[\sum_{i=1}^N a_i\cos(\omega_i x+\alpha_i)\right]^2 }
<1
\ \ \forall x\in {\rm {\bf R}}.
\end{equation}

\noindent
\underbar{\bf Proof:} Define
$\theta_i:=\omega_i x+\alpha_i$, $i\in I$. Then:
\begin{equation}
f(x)\leq {\left[\sum_{i=1}^N
|a_i|\, |\omega_i|\, |\sin(\theta_i)|
\right]^2\over
1-\left[\sum_{i=1}^N |a_i|\, |\cos(\theta_i)|\right]^2 }
<
{\left[\sum_{i=1}^N
|a_i|\, |\sin(\theta_i)|
\right]^2\over
1-\left[\sum_{i=1}^N |a_i|\, |\cos(\theta_i)|\right]^2 } .
\end{equation}
Define the three functions:
\begin{equation}
S(\vec x):=\sum_{i=1}^N |a_i|\sin(x_i),\ \
C(\vec x):=\sum_{i=1}^N |a_i|\cos(x_i),\ \
g(\vec x):={S^2(\vec x)\over 1-C^2(\vec x) },
\end{equation}
where $\vec x:=(x_1,\ldots,x_N)\in
{\rm{\bf R}}^N$.
Since there is always an $\vec x$ such that
$|\sin(\theta_i)|=\sin(x_i)$, $|\cos(\theta_i)|=\cos(x_i)$,
we prove T1 by showing that
$g(\vec x)\leq 1\ \forall \vec x\in {\rm{\bf R}}^N$.
Because of $C^2(\vec x)=$ $\sum_{ij}\, |a_i|\, |a_j|
\cos(x_i) \cos(x_j)\leq$ $\sum_{ij} |a_i|\, |a_j|=$
$(\sum_i |a_i|)^2<1$
$\forall \vec x\in {\rm{\bf R}}^N$,
the function $g$ is well-defined and singularity-free
in ${\rm{\bf R}}^N$.
Since $g$ is
differentiable in ${\rm{\bf R}}^N$ we prove $g\leq 1$
by looking for
the extrema of $g$:
\begin{equation}
{\partial g(\vec x)\over\partial x_k}\ =\ 0\ \ \Rightarrow
(1-C^2(\vec x))S(\vec x)\cos(x_k)-S^2(\vec x)C(\vec x)\sin(x_k)=0,
\ \ k\in I.
\label{T1}
\end{equation}
Let $\vec x^*$ be a solution of
(\ref{T1}).
There are three different cases.
(i) $S(\vec x^*)=0$. In this case we have
$g(\vec x^*)=0<1$.
(ii) $C(\vec x^*)=0$ and  $S(\vec x^*)\neq 0$.
In this case we have
$g(\vec x^*)=S^2(\vec x^*)<1$.
(iii) $C(\vec x^*)\neq 0$ and  $S(\vec x^*)\neq 0$.
In this case (\ref{T1}) reduces to
\begin{equation}
\sin(x^*_k)=
{(1-C^2(\vec x^*))\over S(\vec x^*)C(\vec x^*)}\cos(x^*_k),
\ \ k=\in I.
\label{T2}
\end{equation}
For $g$ evaluated at $\vec x^*$ of (\ref{T2}) we obtain:
\begin{equation}
g(\vec x^*)=\left[{1-C^2(\vec x^*)\over
S(\vec x^*)\, C(\vec x^*)  }\right]^2\
{\left[\sum_{i=1}^N |a_i|\cos(x_i^*)\right]^2 \over
1-C^2(\vec x^*) } \ =\
{1-C^2(\vec x^*) \over
S^2(\vec x^*) } \ =\ {1\over g(\vec x^*)  }.
\end{equation}
This implies $g^2(\vec x^*)=1$, or, since
$g\geq 0$ in ${\rm{\bf R}}^N$,
$g(\vec x^*)=1$.
Since there are no boundaries to consider
where absolute maxima of $g$ might be located,
the local extrema of $g$ encompass all the maxima of
$g$ in ${\rm{\bf R}}^N$ and
we have $g\leq 1$ in ${\rm{\bf R}}^N$.
This proves T1.

\noindent
\underbar{\bf Theorem T2:}
Consider the spectral equation
\begin{equation}
F(x):=\cos(x)-\Phi(x)=0,
\label{S1}
\end{equation}
where
\begin{equation}
\Phi(x)=\sum_{i=1}^N a_i\cos(\omega_i x+\alpha_i)
\label{S2}
\end{equation}
with $a_i$, $\omega_i$,
$\alpha_i, x\in {\rm{\bf R}}$,
$i\in I:=\{1,\ldots,N\}$, $N\in {\rm {\bf N}}$,
$\sum_{i=1}^N |a_i|<1$,
and $|\omega_i|<1$. Then there is
exactly one zero $x_n^*$ of (\ref{S1})
in every open interval $I_n=(\nu_n,\nu_{n+1})$,
$\nu_n=n\pi$,
$n\in {\rm{\bf Z}}$.

\noindent
\underbar{\bf Proof:}
\par\noindent
(i) First we observe that $|\Phi(x)|\leq \sum_{i=1}^N
|a_i|<1\ \forall x\in {\rm{\bf R}}$.
\par\noindent
(ii) We use (i) to verify that the points $\nu_n$
are not roots of (\ref{S1}):
$|F(\nu_n)|=|(-1)^n-\Phi(\nu_n)|\geq 1-|\Phi(\nu_n)|
\ {\buildrel (i)\over >} 0$.
This means that roots of (\ref{S1}) are indeed found only
in the {\it open} intervals $I_n$.
\par\noindent
(iii) We define the closures $\bar I_n=[\nu_n,\nu_{n+1}]$.
In $\bar I_n$ we define $\xi$ according to
\begin{equation}
x=\nu_n+\xi,\ \ \ \ 0 \leq\xi\leq \pi.\ \ \
\label{S3}
\end{equation}
Inserting (\ref{S3}) into (\ref{S1}) we see that
in $\bar I_n$ the spectral function
$F(x)$ is identical with
\begin{equation}
 f_n(\xi)=(-1)^n\cos(\xi) - \varphi_n(\xi),
\label{fnxi}
\end{equation}
where
\begin{equation}
\varphi_n(\xi) = \sum_{i=1}^N a_i
\cos(\omega_i\xi+\alpha_i+n\pi\omega_i).
\end{equation}
\par\noindent
(iv) Because of (i): ${\rm sign}\, F(\nu_n)=(-1)^n$.
We use this fact to show:
${\rm sign}\, F(\nu_n)F(\nu_{n+1})=(-1)^{2n+1}=-1$.
Since $F$ is continuous, this proves that
there is at least one root of $F$ in every $I_n$,
$n\in {\rm{\bf Z}}$.
\par\noindent
(v) According to (iii) and (\ref{fnxi})
the roots of $F$ in $I_n$ satisfy
$(-1)^n\cos(\xi) = \varphi_n(\xi)$, or
\begin{equation}
   \xi=\beta_n(\xi),
\label{S4}
\end{equation}
where
$\beta_n(\xi) = \arccos[(-1)^n \varphi_n(\xi)]$.
Therefore, roots of $F$
are fixed points of $\beta_n$.
\par\noindent
(vi) In $\bar I_n$:
\begin{equation}
[\beta_n'(\xi)]^2 = {\left[ \sum_{i=1}^N a_i \omega_i
\sin(\omega_i\xi+\alpha_i+n\pi\omega_i)\right]^2 \over
1-\left[ \sum_{i=1}^N a_i
\cos(\omega_i\xi+\alpha_i+n\pi\omega_i)\right]^2 }
\ {\buildrel T1\over <}\ 1\ \ \Rightarrow
\beta_n'(\xi)<1.
\end{equation}
(vii) Because of (vi) the conditions for
L16 are fulfilled and $\beta_n$ has at most
one fixed point in $I_n$. This means that
$F$ has at most
one root in $I_n$.
Since according to (iv)
there is at least one root of $F$
in $I_n$, it follows that $F$ has
exactly one root in $I_n$.

\noindent
\underbar{\bf Theorem T3:}
\begin{equation}
\lim_{\epsilon_1,\epsilon_2\rightarrow 0}\,
 \int_{x^*-\epsilon_1}^{x^*+\epsilon_2}\,
\left(\sum_{n=1}^{\infty}\, {e^{in\sigma(x)}\over n}
\right)\, dx\ =\ 0\ =\
\lim_{\epsilon_1,\epsilon_2\rightarrow 0}\,
 \sum_{n=1}^{\infty}
 \left( \int_{x^*-\epsilon_1}^{x^*+\epsilon_2}\,
{e^{in\sigma(x)}\over n}\, dx
\right),
\label{t31}
\end{equation}
where $\epsilon_1,\epsilon_2>0$,
$\sigma(x^*)$ mod $2\pi=0$ and
$\sigma(x)$ is continuous and has a continuous
first derivative.

\noindent
\underbar{\bf Proof:}
\par\noindent
The two limits in (\ref{t31}) are independent. Therefore,
splitting the integration range
into two pieces (allowed with L11), one from
$x^*-\epsilon_1$ to $x^*$, and the other from
$x^*$ to $x^*+\epsilon_2$,
it is enough to prove
\begin{equation}
\lim_{\epsilon\rightarrow 0}\,
 \int_{x^*}^{x^*+\epsilon}\,
\left(\sum_{n=1}^{\infty}\, {e^{in\sigma(x)}\over n}
\right)\, dx\ =\ 0\ =\
\lim_{\epsilon\rightarrow 0}\,
 \sum_{n=1}^{\infty}
 \left( \int_{x^*}^{x^*+\epsilon}\,
{e^{in\sigma(x)}\over n}\, dx
\right),
\label{t32}
\end{equation}
where $\epsilon>0$. The case $\epsilon<0$,
covering the other integral in
(\ref{t31}) is treated in complete analogy.
The first equality in (\ref{t32})
follows immediately from
F2, F3 and the fact that according to L20
the real part has a Riemann-integrable
log singularity at $x=x^*$ and
the imaginary part has a
Riemann-integrable jump-singularity at $x=x^*$.
The second equality is more difficult to prove.
\par
For the following considerations we
assume $\sigma'(x^*)\neq 0$.
We will comment on the case $\sigma'(x^*)=0$ below.
According to L10 $\exists\epsilon^*>0:$
$f(x)=\sigma'(x)/\sigma'(x^*)>1/2$,
for $|x-x^*|<\epsilon^*$. Let $\epsilon<\epsilon^*$:
$$\left| \int_{x^*}^{x^*+\epsilon}\,
{\exp[in\sigma(x)]\over n}\, dx\right|
\ {\buildrel \rm L10 \over <}
\left| \int_{x^*}^{x^*+\epsilon}\,
{\exp[in\sigma(x)]\over n}\, 2{\sigma'(x)\over\sigma'(x^*)}
\, dx\right| = $$
\begin{equation}
  {2\over n^2|\sigma'(x^*)|}\,
\left|\exp[in\sigma(x^*+\epsilon)]-1
\right|\leq {4\over n^2|\sigma'(x^*)|},
\label{T3a}
\end{equation}
where the last estimate is a simple consequence of the fact
that the exponential function is unimodular.
While this simple estimate will be useful later on for the case
of large $n$, we need a better estimate for small $n$:
\begin{equation}
\left|\exp[in\sigma(x^*+\epsilon)]-1
\right|=
\left|\exp\left\{in\epsilon
{[\sigma(x^*+\epsilon)-\sigma(x^*)]\over \epsilon}
\right\}\  -1\, \right|.
\label{t3b}
\end{equation}
Now because $\sigma(x)$ is differentiable,
we have according to the intermediate value theorem
of differential calculus:
$\exists\xi(\epsilon):[\sigma(x^*+\epsilon)-\sigma(x^*)]
/\epsilon=\sigma'(\xi)$; $\xi\in[x^*,x^*+\epsilon]$.
Therefore:
\begin{equation}
 \left|\exp[in\sigma(x^*+\epsilon)]-1
\right|=
\left|\exp[in\sigma'(\xi)\epsilon]-1\right|
\ {\buildrel \rm L6 \over <}
2|n\sigma'(\xi)\epsilon|=2n\epsilon|\sigma'(\xi)|
\label{t3c}
\end{equation}
for $|n\sigma'(\xi)\epsilon|<2$, or,
$n\epsilon|\sigma'(\xi)|/2<1$.
Let
\begin{equation}
 N(\epsilon)=
\left\lfloor {2
\over \epsilon |\sigma'(\xi(\epsilon))|}\right\rfloor
\label{t3d}
\end{equation}
where $\lfloor\ \rfloor$ is the floor function ($\lfloor x\rfloor$:
largest integer smaller than $x$). Then:
\begin{equation}
 \left| \int_{x^*}^{x^*+\epsilon}
{\exp[in\sigma(x)]\over n}\,
dx\right|\leq {4\epsilon\over n}\,
\left|{\sigma'(\xi(\epsilon))\over
\sigma'(x^*)}\right|
\ {\buildrel \rm L10 \over <}\
{6\epsilon\over n}\ \ {\rm for }\  n\leq N(\epsilon).
\label{t3e}
\end{equation}
So now we have
$$
\left|\sum_{n=1}^{\infty}\,
 \int_{x^*}^{x^*+\epsilon} {\exp[in\sigma(x)]\over n} dx
\right|
\leq  $$
$$
\sum_{n=1}^{N(\epsilon)}\, \left|
 \int_{x^*}^{x^*+\epsilon} {\exp[in\sigma(x)]\over n} dx
\right|+
\sum_{n=N(\epsilon)+1}^{\infty}\, \left|
 \int_{x^*}^{x^*+\epsilon} {\exp[in\sigma(x)]\over n} dx
\right|\leq $$
\begin{equation}
 \sum_{n=1}^{N(\epsilon)}\, {6\epsilon\over n}\ +\
\sum_{n=N(\epsilon)+1}^{\infty}{4\over n^2|\sigma'(x^*)|}.
\label{t3f}
\end{equation}
Both sums vanish in the limit of $\epsilon\rightarrow 0$. We
show this in the following way. For the first sum we obtain:
$$ \sum_{n=1}^{N(\epsilon)}\, {6\epsilon\over n}\
\ {\buildrel \rm L9 \over <}\
6\epsilon [1+\ln(N(\epsilon))]=
6\epsilon\left(1+\ln\left\lfloor{1\over \epsilon
|\sigma'(\xi(\epsilon))|}\right\rfloor\right) <$$
$$ 6\epsilon\left\{1-\ln(\epsilon)]-\ln\left|
{\sigma'(\xi(\epsilon))\over
\sigma'(x^*)}\right|-\ln|\sigma'(x^*)|\right\} < $$
\begin{equation}
 6\epsilon\left\{1-\ln(\epsilon)+\ln(2)-\ln|\sigma'(x^*)|
\right\}
\rightarrow 0\ \ {\rm for }\ \epsilon\rightarrow 0.
\label{t3g}
\end{equation}
For the second sum we obtain:
\begin{equation}
\sum_{n=N(\epsilon)+1}^{\infty}\,
{4\over n^2|\sigma'(x^*)|}
\ {\buildrel \rm L8 \over <}\
{4\over |\sigma'(x^*)|N(\epsilon)}<
{12\epsilon\over 4-3\epsilon|\sigma'(\xi(\epsilon))|}
\rightarrow 0\ {\rm for }\ \ \epsilon\rightarrow 0,
\label{t3h}
\end{equation}
where we used
\begin{equation}
N(\epsilon)>{2\over\epsilon |\sigma'(x^*)|}\,
\left|{\sigma'(x^*)\over\sigma'(\xi(\epsilon))}\right|\ -1\
{\buildrel {\rm L10}\over >}\ {4\over 3\epsilon
|\sigma'(x^*)|}\ -1.
\end{equation}

For the above proof we assumed $\sigma'(x^*)\neq 0$.
But our proof still works for the case
$\sigma'(x^*)=0$ if
we use $\sigma''(x)/\sigma''(x^*)$
instead of $\sigma'(x)/\sigma'(x^*)$
in (\ref{T3a}). After a partial integration and noting
that
(i) $\lim_{\epsilon\rightarrow 0}\sigma'(x^*+\epsilon)=0$
and (ii) $\forall |x-x^*|<\epsilon\, \exists\, C(\epsilon)>0,
\,\lim_{\epsilon\rightarrow 0}C(\epsilon)=0:\,
[\sigma'(x)]^2\leq C(\epsilon)|\sigma'(x)|$,
we arrive at an equation very similar to (\ref{T3a}).
Then,
following the steps (\ref{t3b}) -- (\ref{t3h})
establishes T3 in this case too.
This idea can be generalized to the case where the
first non-vanishing derivative is of order $n$, i.e.
$\sigma^{(k)}(x^*)=0$ for $k=0,1,...,n-1$,
$\sigma^{(n)}(x^*)\neq 0$. In this case we use
$\sigma^{(n)}(x)/\sigma^{(n)}(x^*)$ in (\ref{T3a}).
It is not possible that all derivatives of $\sigma(x)$
vanish at $x^*$ since (making the physically justified
assumption that $\sigma(x)$ is an entire function)
this would mean that $\sigma(x)$
is identically zero and
there exists a
scattering channel in which ``nothing happens''.
Since these trivial scattering channels have
no influence on the spectrum of a given quantum graph,
it is possible to
eliminate all trivial scattering channels
and thus reduce
the dimensionality of the $S$ matrix such that none of
the eigenphases of the new,
reduced $S$ matrix is constant.
Taking this into account,
our proof establishes T3 without any exceptions.

\section{Appendix B: Formulae, Definitions and Lemmas}
This appendix is a collection of formulae, definitions and
lemmas needed for the proofs
presented in the text and in Appendix A. They are collected
here since they are lowest in the hierarchy of proof ideas.
``Formulae'' are identities that can be found in
tables. We compiled them
in this appendix for completeness and easy
reference.
``Lemmas'' are simple theorems of a general nature
 that may be found in analysis text books, but are not
usually easily accessible.
So we compiled them here for completeness and convenience.
``Theorems'' are specific to the context of this paper.
They are harder to prove and
unlikely to be found in standard analysis text books.
Every lemma and theorem is proved explicitly,
unless a convenient proof is found in the literature
(see, e.g., L11).

\noindent
\underbar{\bf Formula F1:} $1-\cos(x)=2\sin^2(x/2)$.

\noindent
\underbar{\bf Formula F2:} (see \cite{GR} formula 1.4411, p. 44):
$\sum_{\nu=1}^{\infty}\, {\sin(\nu x)\over \nu}\ =\
  {\pi -x\over 2}, \ \ \ 0<x<2\pi$.

\noindent
\underbar{\bf Formula F3:} (see \cite{GR} formula 1.4412, p. 44):
$\sum_{\nu=1}^{\infty}\, {\cos(\nu x)\over \nu}\ =\
  {1\over 2}\ln\left({1\over 2[1-\cos(x)]}\right)
\ {\buildrel F1\over =}\
  -\ln(2)-\ln[\sin(x/2)],
   \ \ \ 0<x<2\pi$.

\noindent
\underbar{\bf Definition D1:} A function $f$ is differentiable
in $x$ $\iff$ $\exists \lambda\in {\rm {\bf R}} \
\forall \delta >0\
\exists \epsilon^*(\delta)>0\ \forall |\epsilon| < \epsilon^*$:
$$ \left| {f(x+\epsilon)-f(x)\over\epsilon}-\lambda
\right|<\delta .
$$
The constant $\lambda$ is also denoted as $\lambda\equiv
f'(x)$.

\noindent
\underbar{\bf Lemma L1:} Let $a,b\in {\rm{\bf R}}$. Then:
$ |a+ib|=\sqrt{a^2+b^2}\leq |a|+|b|\ (*)$.
\par\noindent
\underbar{Proof:}
$a^2+b^2\leq a^2+2|a|\, |b| + b^2=(|a|+|b|)^2$.
Monotony of the square root yields $(*)$.

\noindent
\underbar{\bf Lemma L2:} $1-x^2/[n(n+1)]>0$ for
$|x|<1$ and $n\in {\rm{\bf N}}$.
\par\noindent
\underbar{Proof:}
$(|x|/n)(|x|/(n+1))<1$ $\Rightarrow$ $1-x^2/[n(n+1)]>0$.

\noindent
\underbar{\bf Lemma L3:} $|\sin(x)|\leq |x|\
\forall x\in {\rm{\bf R}}$.
\par\noindent
\underbar{Proof:} Trivial for $x=0$ and $|x|\geq 1$. Let $|x|<1$:
$|\sin(x)|=\sin(|x|)=|x|-|x|^3[1-x^2/(4\cdot 5)]/3!-
|x|^7[1-x^2/(8\cdot 9)]/7!-\ldots
\ {\buildrel L2\over \leq}\ |x|$.

\noindent
\underbar{\bf Lemma L4:}
$|\sin(x)|\geq |x|/2$ for $|x|<1$.
\par\noindent
\underbar{Proof:} $|\sin(x)|=\sin(|x|)=|x|/2+|x|(1-x^2/3)/2+
|x|^5[1-x^2/(6\cdot 7)]/5!+\ldots
\ {\buildrel L2\over \geq}\ |x|/2$.

\noindent
\underbar{\bf Lemma L5:} $x^2/2\leq |x|$ for $|x|\leq 2$.
\par\noindent
\underbar{Proof:}
$x\geq 0$: $x^2/2-x=x(x-2)/2\leq 0$ for $0\leq x\leq 2$.
\par\noindent
$x< 0$: $x^2/2+x=x(x+2)/2\leq 0$ for $-2\leq x< 0$.

\noindent
\underbar{\bf Lemma L6:} $|\exp(ix)-1|\leq 2|x|$ for $|x|\leq 2$.
\par\noindent
\underbar{Proof:}
$|\exp(ix)-1|=|\cos(x)+i\sin(x)-1|
\ {\buildrel F1\over =}\ |-2\sin^2(x/2)+i\sin(x)|
\ {\buildrel L1\over \leq}\ 2\sin^2(x/2)+|\sin(x)|
\ {\buildrel L3\over \leq}\ 2(x/2)^2+|x|
\ {\buildrel L5\over \leq}\ 2|x|$ for $|x|\leq 2$.

\noindent
\underbar{\bf Lemma L7:} Let $f(x)$ be a monotonically
decreasing function and $f(\nu)=a_{\nu}\in{\rm {\bf R}}$,
$\nu=1,2,\ldots$.
Let $M,N\in{\rm{\bf N}}$, $M<N$.
Then: $\sum_{\nu=M+1}^N\, a_{\nu} \leq\ \int_M^N\, f(x)\, dx$.
\par\noindent
\underbar{Proof:}
According to the intermediate value theorem of
integral calculus $\exists \xi_{\nu}\in [\nu,\nu+1]$ such
that $\int_{\nu}^{\nu+1}\, f(x)\, dx=f(\xi_{\nu})$. Then:
$\int_M^N f(x)dx=\sum_{\nu=M}^{N-1}\int_{\nu}^{\nu+1}
f(x)dx=\sum_{\nu=M}^{N-1}f(\xi_{\nu})\geq
\sum_{\nu=M}^{N-1}a_{\nu+1}=\sum_{M+1}^Na_{\nu}$.

\noindent
\underbar{\bf Lemma L8:} $\sum_{M+1}^{\infty}\,
  {1\over n^{\alpha}}\ \leq\ \int_M^{\infty}\,
  {1\over n^{\alpha}}\, dn\ =\ {1\over (\alpha-1)M^{\alpha-1}}$
for $\alpha>1$.
\par\noindent
\underbar{Proof:}
L7 with $N\rightarrow\infty$.

\noindent
\underbar{\bf Lemma L9:} $\sum_{n=1}^N\,
  {1\over n}\ \leq\ 1+\ln(N)$.
\par\noindent
\underbar{Proof:}
$\sum_{n=1}^N {1\over n}=1+\sum_{n=2}^N{1\over n}
\ {\buildrel L7\over \leq}\ 1+\int_1^N{1\over n}dn=
1+\ln(N)$.

\noindent
\underbar{\bf Lemma L10:} Let $f(x)$ be continuous
in ${\rm{\bf R}}$. Then: $\forall x^*\in{\rm{\bf R}}$
with $f(x^*)\neq 0$
$\exists \epsilon^*(x^*)>0: 1/2 < f(x)/f(x^*) < 3/2\ \ \forall
|x-x^*|<\epsilon^*(x^*)\ \ (\#)$.
\par\noindent
\underbar{Proof:}
Since $f(x)$ is continuous:
$\forall \delta\ \exists\epsilon(\delta):
|f(x)-f(x^*)|<\delta\ \ \forall |x-x^*|<\epsilon(\delta)$.
Choose $\delta=\delta^*:=|f(x^*)|/2$ and define
$\epsilon^*:=\epsilon(\delta^*)$.
Then: $|[f(x)/f(x^*)]-1|
< \delta^*/|f(x^*)|=1/2\ \ \forall |x-x^*|<\epsilon^*$.
This inequality can be used in
two different ways: (i) $1-f(x)/f(x^*)\leq
|[f(x)/f(x^*)]-1|<1/2\ \Rightarrow\
f(x)/f(x*)>1/2\ \
\forall |x-x^*|<\epsilon^*$.
This is the first inequality in (\#).
(ii) $[f(x)/f(x^*)]-1\leq
|[f(x)/f(x^*)]-1|<1/2\ \Rightarrow\
f(x)/f(x^*)<3/2\ \
\forall |x-x^*|<\epsilon^*$.
This is the second inequality in (\#).

\noindent
\underbar{\bf Lemma L11:} Let
$\alpha_{\nu}, \beta_{\nu}\in{\rm{\bf R}}$, $\nu=1,2,\ldots$,
$|\sum_{\nu=1}^{\infty}\alpha_{\nu}|<\infty$,
$|\sum_{\nu=1}^{\infty}\beta_{\nu}|<\infty$,
and $a,b\in{\rm{\bf R}}$. Define:
$S:=\sum_{\nu=1}^{\infty}(a\alpha_{\nu}+b\beta_{\nu})$.
Then: $|S|<\infty$ and
$S=a\left(\sum_{\nu=1}^{\infty}\alpha_{\nu}\right) +
b\left(\sum_{\nu=1}^{\infty}\beta_{\nu}\right)$.
\par\noindent
\underbar{Proof:}
See \cite{EL}, volume II, p. 3, theorem
1.2.1.

\noindent
\underbar{\bf Lemma L12:} $\sum_{\nu=1}^{\infty}\,
\exp(i\nu x)/\nu$ exists and is finite in $0<x<2\pi$.
\par\noindent
\underbar{Proof:} Follows immediately from F2, F3 and L11.

\noindent
\underbar{\bf Lemma L13:}
Let $f(x)$ be differentiable in
$x=a$ with $f(a)=a$ and $f'(a)<1$. Then $\exists c>a$ with
$c-f(c)>0$.
\par\noindent
\underbar{Proof:}
Using D1 as the criterion for differentiability,
we choose
$\lambda=f'(a)<1$,
$\delta=1-\lambda$, $0<\gamma<\epsilon^*(\delta)$ and
$c=a+\gamma$. Then:
$$ \left|{f(a+\gamma)-a\over\gamma}\right|-\lambda \leq
   \left|{f(a+\gamma)-f(a)\over\gamma}-\lambda\right| <
\delta=1-\lambda. $$
$$\Rightarrow\
   \left| {f(a+\gamma)-a\over\gamma}\right| < 1. $$
Now:
$$ c-f(c)=\gamma-{f(a+\gamma)-a\over\gamma}\, \gamma
\geq \gamma-
   \left|{f(a+\gamma)-a\over\gamma}\right|\gamma > 0. $$

\noindent
\underbar{\bf Lemma L14:} Let $f(x)$ be differentiable in
$x=a$ with $f(a)=a$ and $f'(a)<1$. Then $\exists c<a$ with
$f(c)-c>0$.
\par\noindent
\underbar{Proof:}
Analogous to the proof of L13.

\noindent
\underbar{\bf Lemma L15:}
Let $f(x)$ be continuous in $[a,b]$
with $f(a)=a$, $f(b)=b$, $f'(a)<1$ and $f'(b)<1$. Then
there exists at least one additional
fixed point $z$ of $f$ in $[a,b]$ with $a<z<b$.
\par\noindent
\underbar{Proof:}
Because of L13 there exists
$c>a$ with $f(c)<c$. Because of L14 there exists
$d<b$ with $f(d)>d$. Define $H(x)=f(x)-x$. Then
$H(c)<0$ and $H(d)>0$. Since $f$ is continuous, so is $H$.
Then, because of the intermediate value theorem of
calculus, $\exists z, a<c<z<d<b$, with $H(z)=0$, i.e. $f(z)=z$.

\noindent
\underbar{\bf Lemma L16:} Let $f(x)$ be differentiable
in $[a,b]$ with $f'(x)<1\,\forall x\in [a,b]$.
Then $f$ has at most
one fixed point in $[a,b]$.
\par\noindent
\underbar{Proof:}
Assume that $f$ has exactly $n>1$ fixed points
$x_1<x_2<...<x_n$ in $[a,b]$. Then,
because of L15, there must be at least one additional fixed point
between any pair of fixed points $x_j$, $j=1,...,n$,
bringing the total number of fixed points
to at least $2n-1>n$, for $n>1$. This contradicts
the assumption that $f$ has exactly $n>1$ fixed points.
Therefore $f$ cannot have a finite number
$n>1$ of fixed points.
Assume now
that $f$ has a countably infinite
number of fixed points $x_1,x_2,...$
in $[a,b]$. Then, according to Weierstra{\ss},
there exists an accumulation point $x^*$
in $[a,b]$. This implies: $\forall \epsilon > 0\ \exists x_n> x_m$
with $(x_n-x_m)<\epsilon$. Because $f$ is differentiable we have
according to D1:
$$ \left|{f(x_m+\epsilon)-f(x_m)\over\epsilon}-f'(x_m)\right|
<\delta\ \ \ \forall \epsilon<\epsilon^*(\delta).
\eqno(*) $$
Choose $\delta=1-f'(x_m)$, $0<\epsilon=x_n-x_m<\epsilon^*(\delta)$,
and use $f(x_n)=x_n$, $f(x_m)=x_m$, then
$$ \left|{f(x_m+\epsilon)-f(x_m)\over\epsilon}-f'(x_m)\right|=
\left|{f(x_n)-f(x_m)\over x_n-x_m}-f'(x_m)\right|=$$
$$ |1-f'(x_m)|=
1-f'(x_m)=\delta . $$
This contradicts equation (*).
Therefore there cannot be a countably infinite
number of fixed points
in $[a,b]$.
Now assume that $f$ has a continuum of
fixed points
in $[a,b]$.
In this case we can easily show that at an interior point
$x^*$
of the continuum of fixed points we have $f'(x^*)=1$
in contradiction to $f'(x)<1$ $\forall x\in [a,b]$.
Therefore, in summary,
there cannot be any finite number of fixed points $n>1$, nor
can there be
infinitely many fixed points of $f$ in $[a,b]$.
The only alternatives are zero or one fixed point, i.e.
at most one fixed point, as stated in L16.

\noindent
\underbar{\bf Lemma L17:} Let $S$ be a unitary matrix of finite
dimension $B\in{\rm{\bf N}}$, $B\geq 1$. Denote by
$\exp(i\sigma_1)$, $\exp(i\sigma_2)$, $\ldots$, $\exp(i\sigma_B)$
its eigenvalues where $\sigma_j\in {\rm{\bf R}}$ and
$\sigma_j$ mod $2\pi\neq 0$.
Define $M:=\sum_{n=1}^{\infty} S^n/n$. Then:
$|M_{ij}|<\infty$, $i,j=1,\ldots,B$.
\par\noindent
\underbar{Proof:} Since $S$ is unitary, there exists a unitary matrix
$\Omega$ with
$S=\Omega$ ${\rm diag}$ $(\exp(i\sigma_1)$, $\ldots$, $\exp(i\sigma_B))$
$\Omega^{\dagger}$. Also:
$S^n = \Omega {\rm diag} (\exp(in\sigma_1), \ldots, \exp(in\sigma_B))
\Omega^{\dagger}$. Define the series
$\alpha^{(j)}=\sum_{n=1}^{\infty} \exp(in\sigma_j)/n$.
According to L12
these series are convergent. Then,
according to L11, $\mu_{jm}:=\sum_{l=1}^B\Omega_{jl}\alpha^{(l)}
\Omega^*_{ml}$ is convergent, and therefore finite, because only
a finite sum over $\alpha^{(l)}$ is involved. Again with L11:
$\mu_{jm}=\sum_{n=1}^{\infty}(1/n)\sum_{l=1}^B
\Omega_{jl}\exp(in\sigma_l)\Omega_{ml}^*=\sum_{n=1}^{\infty}
(1/n)(S^n)_{jm}=M_{jm}$.
This means that since $\mu_{ij}$ is finite, so is
$M_{ij}$, i.e. $|M_{ij}|<\infty$.

\noindent
\underbar{\bf Lemma L18:} Let $S$ be the matrix of L17. Then:
for $0<\sigma_j<2\pi$, $j=1,\ldots,B$:
${\rm Tr}\sum_{n=1}^{\infty}S^n/n=
\sum_{j=1}^B\{-\ln(2)-\ln[\sin(\sigma_j/2)]+i(\pi-\sigma_j)/2\}$.
\par\noindent
\underbar{Proof:} Because of L11,
$$ {\rm Tr} \sum_{n=1}^{\infty} {1\over n}S^n=
\sum_{n=1}^{\infty}{1\over n}{\rm Tr} S^n=
\sum_{n=1}^{\infty}{1\over n}{\rm Tr} \{\Omega\, {\rm diag}
[\exp(in\sigma_1),\ldots,\exp(in\sigma_B)]\,\Omega^{\dagger}\}= $$
$$ \sum_{n=1}^{\infty}{1\over n}[\exp(in\sigma_1+\ldots +
\exp(in\sigma_B)]
\ {\buildrel {\rm F2,F3,L11}\over =}\
\sum_{j=1}^B\{-\ln(2)-\ln[\sin(\sigma_j/2)]+i(\pi-\sigma_j)/2\}.$$

\noindent
\underbar{\bf Lemma L19:} ${\rm Im Tr}\, \sum_{n=1}^{\infty}
S^n/n=
\sum_{j=1}^B\, (\pi-\sigma_j)/2$.
\par\noindent
\underbar{Proof:} Follows immediately from L18.

\noindent
\underbar{\bf Lemma L20:}
$\ln|\sin(x)|$ has an integrable singularity at $x=0$.
\par\noindent
\underbar{Proof:} Let $b<\pi/2$. Then:
$$ \left|\int_0^b\ln[\sin(x)]dx\right|
\ {\buildrel {\rm L4}\over \leq}\
\left|\int_0^b\ln(x/2)dx\right|=
b\left|\ln(b/2)-1\right| < \infty . $$

\pagebreak

\centerline{\bf Figure Captions}

\bigskip\noindent
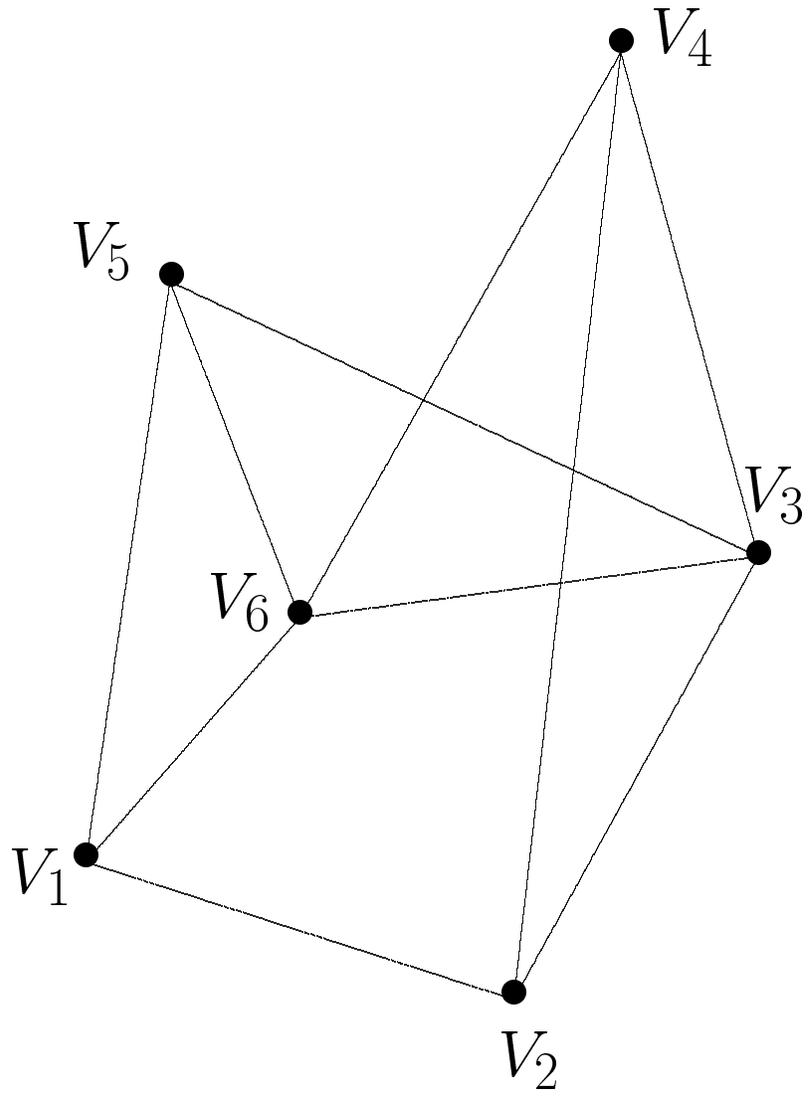
{\bf Fig.~1:} Sketch of a quantum graph with six vertices
              and ten bonds.

\bigskip\noindent
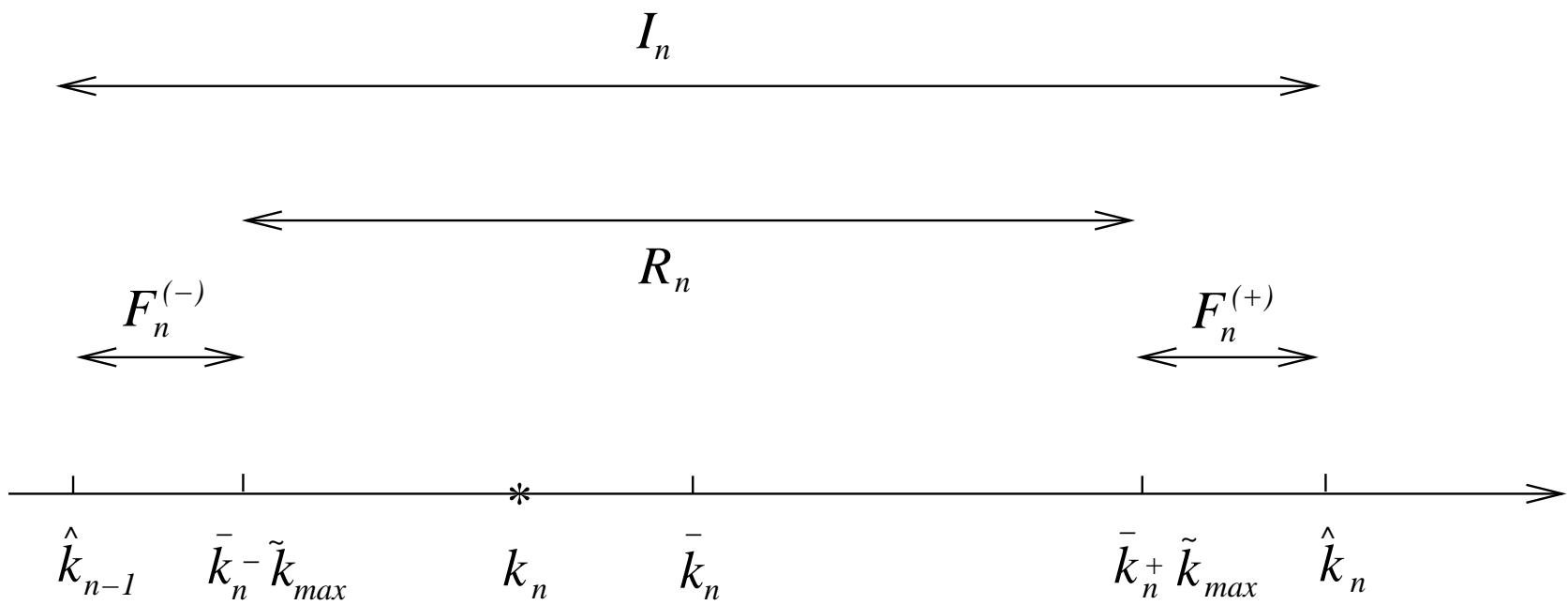
{\bf Fig.~2:} Structure of root cell $I_n$. To the left
              and to the right of $I_n$ are the
              root-free intervals $F_n^{(-)}$ and
              $F_n^{(+)}$, respectively. Together they
              form the root-free zone
              $F_n=F_n^{(-)}\cup F_n^{(+)}$.
              Roots of the spectral equation are found in
              the interval $R_n$. The delimiters of $I_n$
              are $\hat k_{n-1}$ and $\hat k_n$. The
              average location (star) of the root
              $k_n$ is given by $\bar k_n$.

\bigskip\noindent
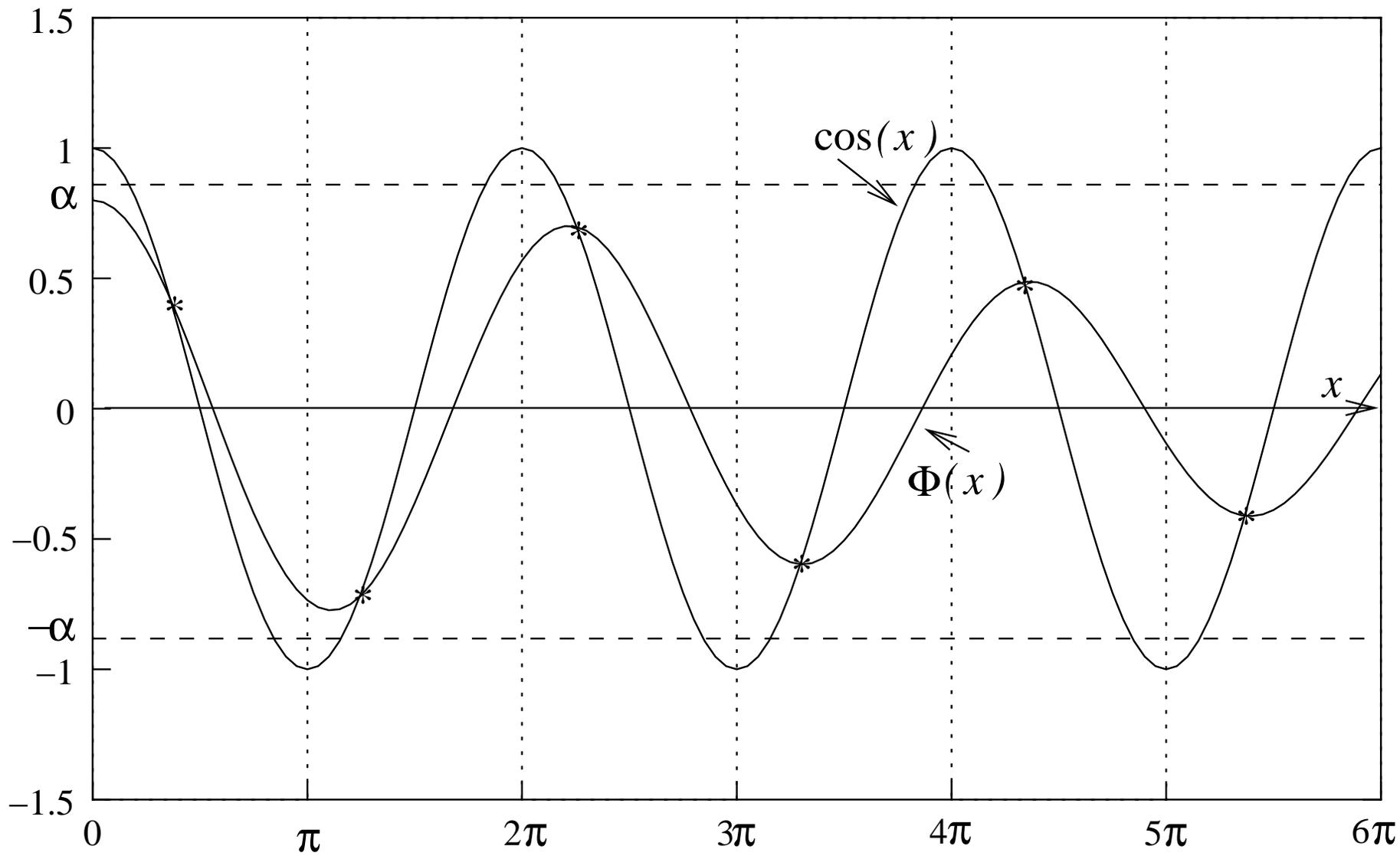
{\bf Fig.~3:} Graphical solution of the spectral equation
              $\cos(x)=\Phi(x)$. Since $\cos(x)$ is
              ``faster'' than $\Phi(x)$, one and only one
              solution exists in every interval
              $(n\pi,(n+1)\pi)$. This fact is proved
              rigorously in Sec.~III.

\bigskip\noindent
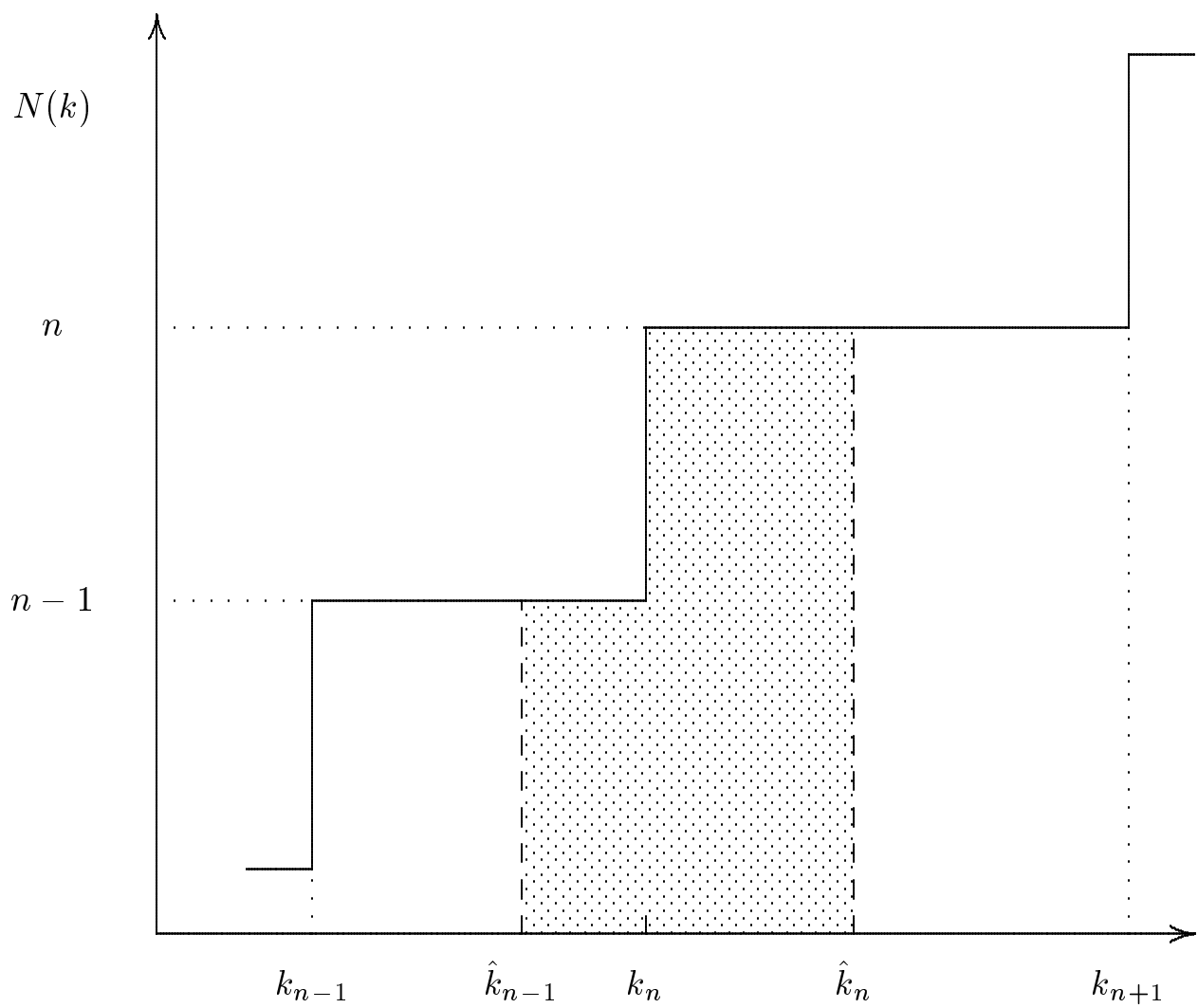
{\bf Fig.~4:} Detail of the spectral staircase
              illustrating the computation of
              the integral (\ref{Nint}) over the spectral
              staircase from $\hat k_{n-1}$ to $k_n$
              for the purpose of obtaining an explicit
              expression for $k_n$.

\bigskip\noindent
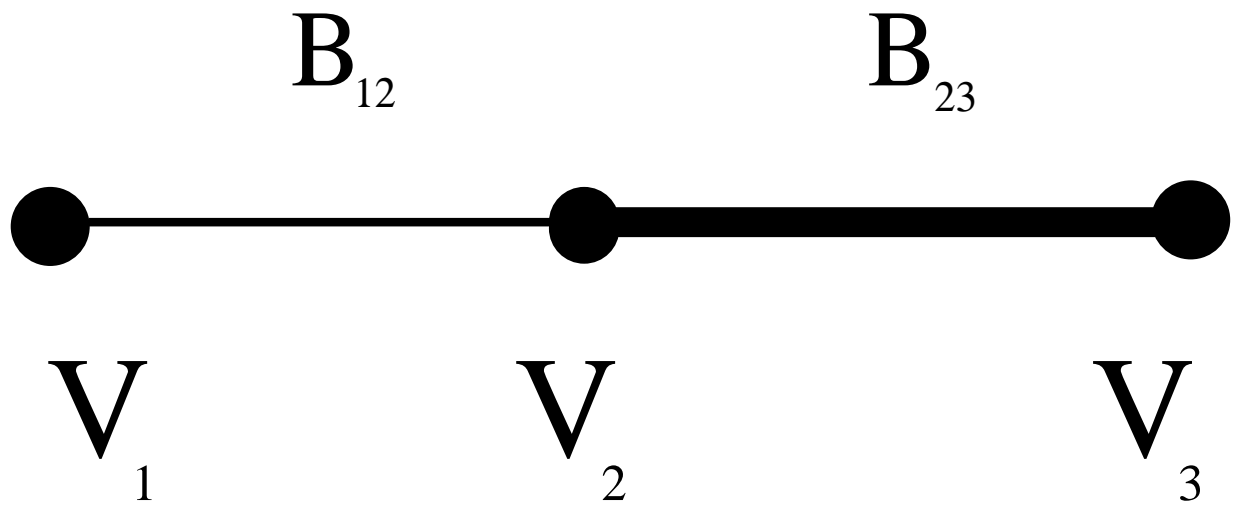
{\bf Fig.~5:} Three-vertex linear quantum graph.
              The vertices are denoted by $V_1$, $V_2$
              and $V_3$, respectively.
              $V_1$ and $V_3$ are
              ``dead-end'' vertices. The bonds are
              denoted by $B_{12}$ and $B_{23}$, respectively.
              While there is no potential on the bond
              $B_{12}$ (indicated by the thin line representing
              the potential-free bond $B_{12}$),
              the bond $B_{23}$ is ``dressed''
              with the energy-dependent
              scaling potential $U_{23}=\lambda E$ (indicated by
              the heavy line representing the ``dressed''
              bond $B_{23}$).

\end{document}